\documentclass[reprint,superscriptaddress,showpacs,amsmath,amssymb,aps,prb,floatfix]{revtex4-2}
\usepackage{subfiles}   

\usepackage{dsfont}
\usepackage{graphicx}
\usepackage{subfigure}
\usepackage[colorlinks=true,allcolors=blue]{hyperref}
\usepackage{xr}
\usepackage[normalem]{ulem}
\usepackage{accsupp}

\usepackage{placeins}
\usepackage[dvipsnames]{xcolor}
\usepackage{xr}

\setcitestyle{super}

\newcommand{\etal}{\emph{et al.}}
\newcommand{\inlinecite}[1]{{\setcitestyle{numbers}\citep{#1}}}
\newcommand{\hS}{\hat{S}}
\newcommand{\PI}{\text{PI}}

\newcommand{\resetpages}{%
  \clearpage
  \setcounter{page}{1}
  \pagenumbering{arabic}%
}
\begin{document}


\setlength{\fboxsep}{0pt}
\setlength{\fboxrule}{.1pt}

\title{Reconstruction of spin structures from topological charge distributions via generative neural network systems}


\author{Kyra H. M. Klos}
\email{kyklos@uni-mainz.de}
\affiliation{%
 Institute of Physics, Johannes Gutenberg-University Mainz, 55128 Mainz, Germany
}%

\author{Jan Disselhoff}
\affiliation{%
 Institute of Computer Science, Johannes Gutenberg-University Mainz, 55128 Mainz, Germany
}%

\author{Michael Wand}
\affiliation{%
 Institute of Computer Science, Johannes Gutenberg-University Mainz, 55128 Mainz, Germany
}%

\author{Karin Everschor-Sitte}%
\affiliation{%
 Faculty of Physics and Center for Nanointegration Duisburg-Essen (CENIDE), University of Duisburg-Essen, 47057 Duisburg, Germany
}%

\author{Friederike Schmid}
\email{schmidfr@uni-mainz.de}
\affiliation{%
 Institute of Physics, Johannes Gutenberg-University Mainz, 55128 Mainz, Germany
}%

\date{\today}

\begin{abstract}

Localized topological defects inherently possess a multiscale
character. While their microstructure configuration depends on the
specific physical system, their topological features and mutual
interactions can be described on the macroscale in terms of a particle
representation.  However, determining the physical properties
associated with a given defect pattern often requires knowledge of the
underlying microscopic structure. In this work, we extend a
Wasserstein generative adversarial neural network by incorporating
physical constraints and Fourier-space information to generate
microscopic spin configurations consistent with prescribed macroscopic
patterns and thermodynamic parameters.  Using the two-dimensional XY
model as a test case, where vortex–antivortex pairs act as long-range
interacting defects, we show that the model generates spin
configurations that accurately reproduce magnetization,
susceptibility, helicity modulus, and spin–spin correlations over a
wide range of temperatures below the Kosterlitz–Thouless transition.
At the same time, deviations in the specific heat reveal limitations
in reproducing higher-order energy fluctuations. A complementary
analysis based on topological data analysis uncovers subtle
differences in global spin-correlation structures at near-critical
temperatures that are not apparent from conventional correlation
functions alone. These results demonstrate both the promise and
current limitations of generative approaches for multiscale studies of
defect-dominated spin systems and at the same time highlight topological 
methods as valuable tools for characterizing critical behavior.

\end{abstract}

\maketitle
\section{Introduction}
\label{sec:intro}

Physics-based descriptions of natural processes often rely on
effective theories. Rather than attempting a full-scale description of
a phenomenon, such theories focus on the dominant processes at a given
scale in the context of the surrounding scales~\cite{Georgi:1993}.
Within this framework, nontrivial topological structures pose
particular challenges due to their intrinsic multiscale nature and
their typically long lifetimes. Topological point defects, for
example, arise as localized perturbations of an underlying ordering
field. While their structure is defined microscopically, they can
often be treated as effective quasiparticles at larger length scales.
Such defects occur in a wide range of materials and systems, including
magnetic thin films~\cite{Vasudevan_td_thinfilms_experimental,
Muhlbauer_skyrmions_in_thin_film}, liquid
crystals~\cite{Kleman_TD_NLC,Darmon_td_liquid_crystal_experimental},
and active matter ~\cite{bacteria_transport,Shankar_TD_active_matter}.
Their intrinsic stability makes them attractive for a range of
applications such as data storage, transport, and unconventional
computing~\cite{perspective_skyrmion_computing}. 

Although defect interactions originate from microscopic distortions of
the underlying fields, they can nevertheless strongly influence the
global properties and dynamical processes of a system.  A paradigmatic
example of this multiscale interplay is the
Berezinskii-Kosterlitz-Thouless (BKT) phase transition~\cite{Kost},
where effective interactions between localized topological defects
drive a transition from a quasi-ordered to a fully disordered state in
two-dimensional spin systems.  Comprehensive numerical studies of such
emerging cooperative phenomena, which are ultimately caused by
interfering defect-induced deformation fields, require large-scale
simulations with microscopic resolution and are therefore
computationally demanding. Experimentally, the situation is
even more challenging: It is often difficult to simultaneously access
mesoscale information on defect configurations and microscale
information on the local microscopic structure. Therefore, dynamical
models of defects in materials often operate solely at a mesoscopic
``defect particle'' level and do not explicitly resolve the underlying
microscopic
processes\cite{Harth_active_passive_liquid_crystal_review,
Tang_dynamics_active_passive_liquid_crystal,Brems_simulation_skyrmions,
Thiele_defect_dynamics}. However, microscopic information is essential
for evaluating the impact of defects on the properties of a material.

In recent years, deep learning methods have emerged as powerful tools
for detecting and characterizing topological defects in both
experimental\cite{Minor_CNN_LC, Ren_CNN_VT_detection} and simulated
data\cite{Beach}, as well as for more indirect analyses that infer
defect properties through their impact on the ordering field and the
resulting material
properties\cite{Walters_ML_detection_td_liquid_crystal,
Li_resnet_directorfield, Rodrigues2021deeper, Horenko2021scalable}. 
Beyond defect identification, supervised and
semi-supervised machine learning approaches have been explored to
learn thermodynamic features and phase behavior in systems containing
topological defects.  Methods combining variational autoencoders
(VAEs) with Gaussian sampling~\cite{Cristoforetti_GVAE_XY} or
temperature labeling~\cite{Naravane_cVAE_XY} have demonstrated the
ability to encode phase-specific information and reconstruct selected
thermodynamic observables. Energy-based models, such as restricted
Boltzmann machines, have also shown success in generating physically
realistic spin configurations, albeit primarily in high-temperature
regimes\cite{Zhang_Boltzmann_XY}.  Park \etal\cite{magnetic_structure}
investigated the use of generative network systems to reconstruct 
``plausible'' microscopic spin configurations in a chiral Heisenberg 
model featuring local stripe formation, classifying configurations
containing topological defects as ``implausible''.

In the present paper, we propose a complementary strategy that
explicitly targets systems containing topological defects.  We train
generative neural network systems to reconstruct physically realistic
microscopic spin configurations that are consistent with prescribed
defect distributions.  As a test case, we consider two-dimensional
spin systems with in-plane spin alignment (planar XY model).  A
central challenge is that the network must learn the underlying
topology, which is known to be difficult for generative networks.
While conceptually related to the approach of Park 
\etal\cite{magnetic_structure}, our method goes beyond plausibility
classification and aims to generate ensembles of spin configurations
that are both topologically consistent and distributed according to
the correct Boltzmann statistics.

By bridging macroscopic defect information and microscopic spin
configurations, our approach enables deeper insights into defect
characteristics, interaction patterns, and correlations with the
surrounding spins.  The resulting backmapping model can serve as a
component of multiscale simulation strategies for large spin systems,
where long-time dynamics are described  by coarse-grained
defect-particle models and microscopic configurations are reconstructed 
as needed for further analysis.  It may also assist the interpretation of
experimental measurements by providing likely microscopic realizations
consistent with observed defect structures. More broadly, this work
serves as a proof of concept for the capabilities and limitations of
generative models in learning complex, implicitly defined topological
features from data.

The paper is organized as follows: We first introduce the model system and validation descriptors. We then describe the generative architectures and training procedure, followed by a presentation and discussion of the results. We conclude with a summary and outlook.

\section{Model System}
\label{sec:xy_model}

The classical two-dimensional XY-model on a planar square lattice
describes a system of interacting two-dimensional spin objects
$\hS_i =  (\cos(\vartheta_i),\sin(\vartheta_i)) $ (i.e.,
$|\hS_i |=1$), where $\vartheta_i$ denotes the angle between
$\hS_i$ and an arbitrary axis\cite{KT,Berezinskii, NELSON1979255}.
It is used, among other, to model planar superconductivity as in
Josephson junction arrays or superfluidity in $\text{He}^4$
films\cite{KT,Berezinskii, NELSON1979255}. The Hamiltonian of the XY
model is given by

\begin{equation}
    \label{equ:H_xy}
    E = -J \sum_{<i,j>} \hS_i\cdot\hS_j = -J \sum_{<i,j>} \cos(\vartheta_i - \vartheta_j),
\end{equation}
where
$\langle i,j \rangle$ denotes the sum over all next-neighbor pairs on 
the lattice. In the present work, we consider a ferromagnetic coupling 
with $J>0$, which favors parallel alignment of the spins.
 
 \subsection{Thermodynamic properties}
 \label{subsec:thermo_prop}

For future reference, we briefly recall the essential behavior of the
ferromagnetic XY-model\cite{chaikin_lubensky_1995}.
We assume periodic boundary conditions in both planar directions.

The Hamiltonian \eqref{equ:H_xy} is invariant under the global gauge
transformation $\vartheta \rightarrow \vartheta + \alpha$,
corresponding to a continuous  $O(2)/U(1)$ symmetry. At temperature
zero, the system assumes a symmetry-breaking minimum energy state with
fully aligned spins.  Since the direction of alignment is arbitrary,
this ground state is continuously degenerated.  As a result, there
exist two types of basic excitations: Delocalized spin waves whose
energy vanishes at infinite wave length (massless Goldstone bosons),
and localized topological defects (vortices) with finite energy.  The
latter can be described as distortions of an underlying spin ordering
field surrounding a disordered point-like core. In contrast to spin
waves, vortices cannot be removed by cooperative continuous
deformations of the spin field and therefore have long lifetimes. 

In the planar XY model, the interplay of these two types of
excitations, spin waves and vortices, leads to a particularly
intriguing phase behavior: According to the Mermin-Wagner
theorem\cite{MerWag}, spin waves destroy the long range order at
nonzero temperatures in two dimensions.  At sufficiently low
temperatures, the system nevertheless exhibits so-called quasi long
range order (QLRO), characterized by the fact that the spin-spin
correlation function decays algebraically. At high temperatures, the
QLRO gets lost and the correlation function decays exponentially. The
transition between these two regimes, the
Berezinskii-Kosterlitz-Thouless (BKT) phase transition\cite{KT,
Berezinskii}, is associated with a change in the spatial arrangement
of vortices: At low temperatures, the number of vortices is small, and
they are tightly bound as pairs, stabilized by long-range
vortex-vortex interactions that can be associated with the massless
character of the Goldstone bosons. At the BKT transition, the vortices
unbind and form a free gas, thereby destroying the QLRO. This
transition cannot be described in terms of the standard Landau
framework of phase transitions\cite{landau} and has several
interesting properties, e.g., it is asso\-ciated with an essential
singularity in the free energy of the system.  

\begin{figure}[tb]
    \centering
    \includegraphics[width=.99\columnwidth]{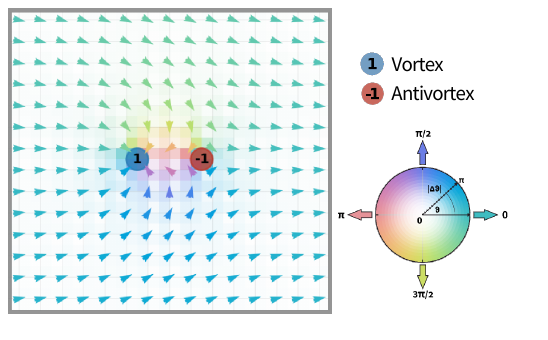}
    \caption{Example of a vortex-antivortex pair (blue and red
    marker) with distance $d_{\nu} = 3$ in a two dimensional XY-model 
    on a square lattice at zero temperature.
    The arrows point in the direction of the spins 
    and are colored according to their orientation as indicated in 
    the color wheel legend (right). The background encodes the local 
    gradient field $\vartheta$. It is colored according to the 
    direction of the gradient, and the saturation level indicates 
    its strength.
    \label{fig:vortex_pair}}
\end{figure}

At sufficiently low temperatures, the orientations of spins vary only
slowly in space except in the core regions of topological defects. In
such cases, the spin configurations can be described by a spin field
$\hS(\mathbf{r})$, or, equivalently, an angle field
$\vartheta(\mathbf{r})$. After expanding the cosine term in the
Hamiltonian \eqref{equ:H_xy} up to second order, $\cos(\vartheta_i -
\vartheta_j) \approx 1 - \frac{1}{2}( \vartheta_i-\vartheta_j)^2$ and
taking the continuum limit $\sum_i \to \int \text{d}^2 r$, one
can rewrite \eqref{equ:H_xy} in the approximate form 
\begin{equation}
\label{equ:H_xy_continuous} E_c = \mbox{const.} +
\frac{J}{2} \int \;\text{d}^2 r \: (\nabla \vartheta )^2,
\end{equation} 
outside of the cores of topological defects. The
defects can be viewed as ``holes'' in the spatial integration domain.
They are characterized by their topological winding number $k$, also
called topological charge, which is calculated through a contour
integral over a closed path surrounding the defect core 
\begin{equation}
\label{equ:vortex_int} 
\oint d\vartheta =  2 \pi k 
\end{equation} 
with $k \in \mathds{Z}$. If the contour integral surrounds several
defects, their winding numbers add up, giving $\oint d\vartheta =  2
\pi \sum_{j} k_j$. Since the free energy associated with a defect is
found to scale with $k^2$, defects have typically winding numbers $k
=1$ (vortices) or $k=-1$ (antivortices). Moreover, the total sum of
winding numbers in a system with periodic boundary conditions must be
zero, therefore the numbers of vortices and antivortices must be
equal.  An example of a system containing a vortex-antivortex pair can
be seen in Fig.\ \ref{fig:vortex_pair} for the zero temperature case.

In the planar XY model, the BKT transition is encountered
at the reduced temperature\cite{Hasenbusch2005XY} 
$k_B T/J = 0.8929(1)$ (with the Boltzmann constant $k_B$).
A signature of the transition is a change in the so-called
helicity modulus or spin-wave-stiffness, which gives the change of the
free energy in response to an imposed macroscopic gradient $\alpha$ of
the mean spin orientation along a direction $\hat{e}$. Such a gradient
can be imposed, e.g., through the boundary conditions. The helicity
modulus is defined as\cite{Fischer_HM} $\Upsilon = \frac{1}{L^2}\frac{\partial^2
F}{\partial \alpha^2}|_{\alpha = 0}$ ($L^2$ is the
system size) and can be measured in equilibrium simulations
with periodic boundary conditions via the expression: 
\begin{equation}
    \label{equ:helicity_modulus}
    \begin{split}
    \Upsilon &= \frac{1}{L^2} \left\langle J 
      \sum_{<i,j>}\cos(\vartheta_i - \vartheta_j) 
      \: (\hat{e}\cdot\hat{\epsilon}_{ij})^2 \right\rangle\\
    &- \frac{1}{k_BTL^2} \left\langle \bigg(J \sum_{<i,j>}
      \sin(\vartheta_i - \vartheta_j) \: (\hat{e} 
      \cdot \hat{\epsilon}_{ij})\bigg)^2  \right\rangle
    \end{split}
\end{equation}
with $\hat{\epsilon}_{ij}$ denoting the bond between spin side $i$ and
$j$ and $\langle\dots\rangle $ the thermal average.  Below the BKT
transition, in the presence of quasi long range order, the system
 resists to   global spin gradients, resulting in a nonzero
helicity modulus. Above the transition, the helicity modulus drops to
zero.

Apart from the helicity modulus, we will also consider other response
functions such as the magnetic susceptibility, which describes the
response of the magnetization, $\mathbf{M} = \sum_i \hS_i$, to an
external magnetic field, and can be measured in the field-free systems
from the fluctuations of the magnetization,
\begin{equation}
    \label{equ:magnetic_susceptibility}
    \chi = \frac{1}{k_B T} \left(\langle 
      \mathbf{M}^2\rangle-\langle \mathbf{M}\rangle^2 \right).
\end{equation}
and the specific heat, which can be measured via
\begin{equation}
    \label{equ:specific_heat}
    \hskip-0.3cm C 
      = \frac{1}{k_B T^2}\left(\langle 
      E^2 \rangle - \langle E \rangle^2 \right)
\end{equation}
Other quantities of interest are the Binder cumulant\cite{binder}
\begin{equation}
    \label{equ:binder}
    U = 1- \frac{\langle (\mathbf{M}^2)^2\rangle}
    {3\langle \mathbf{M}^2 \rangle^2}
\end{equation}
and the thermally averaged magnetization per spin, 
$\langle |\mathbf{M}|/N\rangle$. Even though this latter quantity 
is predicted to vanish in the thermodynamic limit due to the 
Mermin-Wagner theorem, it can still be used to quantify the behavior 
of finite systems as a function of temperature. 
Finally, to analyze local structure, 
we will evaluate the correlation function between two spins,
\begin{equation}
\langle \hS_i \hS_j \rangle = 
\langle \cos(\vartheta_i-\vartheta_j) \rangle
\label{equ:corr}
\end{equation}
as a function of their spatial distance $d_{ij}$.

\subsection{Simulation Details}

\label{subsec:simulation}

To generate training data for our reconstruction method, we 
performed Monte Carlo simulations of the two-dimensional XY model on a
quadratic spin lattice with $N$ spins on a square simulation box with
periodic boundaries in both directions. Unless stated otherwise,
the system size is $16 \times 16$.  Our training data are
spin configurations corresponding to given specified defect
configurations (i.e., distribution of defect positions and charges).
Therefore, we impose fixed defect configurations as hard constraints
in the simulations. Specifically, a defect configuration $\{k_r\}$ is
quantified {\em via} a discretized version of equation
(\ref{equ:vortex_int}) as follows: For each plaquette $\Box_r$ of the
lattice, we define a winding number  
\begin{equation}
    \label{equ:vortex_discrete}
    k_r = \frac{1}{2\pi}\sum_{\{i,j\} \in \Box_{r}} 
    \text{saw}\left(\vartheta_j-\vartheta_i\right)
\end{equation}
where the sum runs over the four pairs $(i,j)$ of spins surrounding
the plaquette in counter-clockwise direction, and 
$\text{saw}(\Delta \varphi) = \text{Arg}(\text{e}^{i\Delta \varphi}) 
\in [-\pi,\pi]$ 
denotes the principal value of the phase difference, mapping angles onto
the interval $[-\pi,\pi]$. Note that the $k_r$ are integer numbers.
In the simulations, we use as initial state the continuous
zero-temperature solution for given defect configuration $\{k_r\}$ to
speed up the equilibration, and then update the spins using a
Metropolis criterion with the constraint that the set $\{k_r\}$ may
not change. 

\section{Topological Data Analysis}
\label{sec:TDA}

The spin configurations obtained from simulations (``real'' data) and
the ones produced by our generative model (generated or ``fake'' data)
are characterized based on a set of descriptors, such that they can be
compared with each other at a quantitative level. These include the
physical observables introduced earlier, such as the thermally
averaged energy, helicity modulus etc. However, physical observables
are typically expressed as statistical averages of local quantities or
pair correlation functions, and do not capture more complex, global
features of the configurations. Therefore, we complement them
with descriptors based on a topological
analysis of the spin configurations. 
These provide complementary integrated information on the entire
hierarchy of $n$-body correlations in the system, and on connectivity
structures across all scales, which is particularly valuable in the
presence of topological defects.

The topology $T$ of a system $X$ is characterized by its k-dimensional
homology groups $\mathcal{H}_k(X)$, which encode information about the
structure of the space and are invariant under homeomorphisms. A
simple way to quantify the homology groups $\mathcal{H}_k(X)$ is via
the Betti-numbers, $\beta_k = \dim \mathcal{H}_k(X)$. Intuitively,
$\beta_k$ counts the number of independent $k$-dimensional holes in
$X$ for $k >0$. The zeroth Betti-number, $\beta_0$, corresponds to the
number of connected components. For mathematical details, we refer 
to Ref.\inlinecite{Nakahara}. The application of these concepts to spin
systems will be explained below.

To capture more detailed and physically meaningful multiscale features 
of spin configurations, we go beyond Betti numbers and use persistent
homology\cite{persistent_homology}. Persistent homology not only
measures which topological features exist but also tracks how long
they persist across different scales. The basic idea is to transform a
spin lattice into a simplicial complex -- a structure built from
points (vertices), line segments (edges), and higher-dimensional faces
(plaquettes) -- and then define a filtration, which is a sequence of
nested subcomplexes. Each step of the filtration adds more elements to
the subcomplex based on a specific rule, and the topological features
that appear and disappear along the filtration are recorded.

\begin{figure}[t]
    \centering
    \includegraphics[width=0.95\columnwidth]{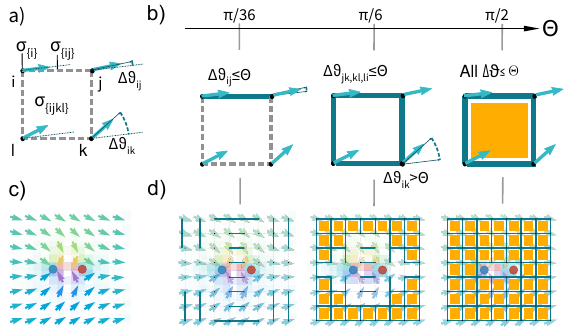}
    \caption{ 
    Construction of cubical complexes $S_\Theta$ 
    for configurations of the XY model on a square lattice
    (Eq.~\ref{equ:filtration}). 
    a) Single square made of lattice sites $i,j,k,l$
    showing basic units $\sigma_{\{i\}}$ (vertices), 
    $\sigma_{\{ij\}}$ (edges), $\sigma_{\{ijkl\}}$ (plaquettes),
    and a local spin configuration (cyan arrows).
    $\Delta \vartheta_{nm}$ denotes the angle between spins on 
    lattice sites $n$ and $m$.
    b) Filtration process for the configuration in a) for three
    filtration threshold values $\Theta = \pi/36, \pi/6, \pi/2$ 
    (left to right). Edges are part of $S_\Theta$
    (marked as thick teal lines) if the angle $\Delta \vartheta$ 
    between their adjacent spins does not exceed the threshold 
    $\Theta$. A plaquette is part of $S_\Theta$ (marked as orange square)
    if none of the angles between any two associated spins exceeds 
    $\Theta$.  For $\Theta=\pi/36$, only the edge $\sigma_{\{ij\}}$ 
    satisfies the filtration criterion. For $\Theta=\pi/6$,
    all edges satisfy the criterion, but the plaquette does not,
    since $\Delta \vartheta_{ik} > \Theta$. 
    For $\Theta = \pi/2$, the plaquette is also part of $S_\Theta$.
    c,d) Illustration of the filtration process for a specific
    spin configuration with one vortex-antivortex pair (blue and 
    red dot) at zero temperature:
    c) Underlying spin configuration (same color coding as in 
    Fig.\ \protect\ref{fig:vortex_pair}).
    d) Resulting filtered cubical complex for the
    filtration thresholds $\Theta=\pi/36, \pi/6, \pi/2$ 
    (from left to right).  As in b), edges in $S_\Theta$ 
    are marked as teal lines, and plaquettes in $S_\Theta$ 
    as orange squares.
    \label{fig:filtration_example}}
\end{figure}

For our spin lattices, we follow an approach proposed by N. Sale
\etal\cite{XY_filtration}, in which the spin lattice is mapped
onto a cubical complex $\Sigma$. The elements of the complex, are the
vertices, edges (nearest-neighbor bonds), and plaquettes. We define a
filtration function $f: \Sigma \to \rm I\!R $  based on the
unsigned angle $\Delta \vartheta_{ij}$ between spins
at sites $i$ and $j$:
\begin{align}
\begin{split}
     \label{equ:filtration}
     \text{Vertex }(V) &: f(\sigma_\{i\}) = 0   \\
     \text{Edge }(E) &: f(\sigma_{\{ij\}}) 
        = \frac{1}{2\pi} \Delta \vartheta_{ij} \\
     \text{Plaquette }(P) &: f(\sigma_{\{ijkl\}})
        = \frac{1}{2\pi}\max_{\{n,m\} \in \{i,j,k,l\}} 
            \Delta \vartheta_{nm}.
 \end{split}
 \end{align}
Here, $\{ij\}$ and $\{ijkl\}$ denote nearest-neighbor pairs and
plaquette sites, respectively. A specific 
filtered complex ${\cal S}_\Theta$ is obtained by choosing a
threshold value $\Theta \ge 0$ and including all elements $\sigma \in
\Sigma$ with $f(\sigma) \le \Theta$.  By construction, all vertices
are always included.  As $\Theta$ increases, edges and plaquettes are
progressively added, with the property that an element can only be
added if all of its boundary elements are either already present or
also being added, consistent with the definition of a simplicial
complex; for example, $P \in {\cal S}_\Theta$ implies $E \in {\cal
S}_\Theta$ for all $E \in \partial P$.  At $\Theta=\pi$, all elements
are included, and ${\cal S}_{\pi}=\Sigma$.
Fig.~\ref{fig:filtration_example} illustrates this process.

Having specified the filtration function, we can analyze the
persistence of the homology groups ${\cal H}_0$ and ${\cal H}_1$ for a
given spin lattice configuration.  To this end, we number the elements
of $\Sigma$ such that $\alpha < \beta$ whenever $f(\sigma_\alpha) <
f(\sigma_\beta)$ and/or $\sigma_\alpha \in \partial \sigma_\beta$. We
then construct a sequence of cubical complexes, ${\cal
S}_\alpha = \bigcup_{\beta=1}^\alpha \sigma_\beta$, adding elements
one by one, and track the evolution of the resulting topological
structure. 
This procedure enables a unique labeling of the homology classes in
${\cal H}_p$ in terms of their so-called ``creators'', i.e., the
building blocks $\sigma_\alpha$ whose inclusion causes the
corresponding homology class to appear for the first time. Conversely,
the addition of a new building block $\sigma_\beta$ may destroy a
homology class or merge two existing classes, thereby eliminating one
of them -- by definition the one that was created last. In this case
$\sigma_\beta$ is referred to as ``destructor'' and is paired with the
creator $\sigma_\alpha$ of the homology class that is eliminated. For
${\cal H}_0$, the homology classes correspond to connected components:
their creators are vertices, while their destructors are edges that
merge two components. For ${\cal H}_1$, the homology classes are
non-boundary cycles: their creators are edges that close such cycles,
and their destructors are plaquettes whose inclusion turns a cycle
into a boundary. 

\begin{figure}[tb]
    \centering
    \includegraphics[width=0.85\columnwidth]{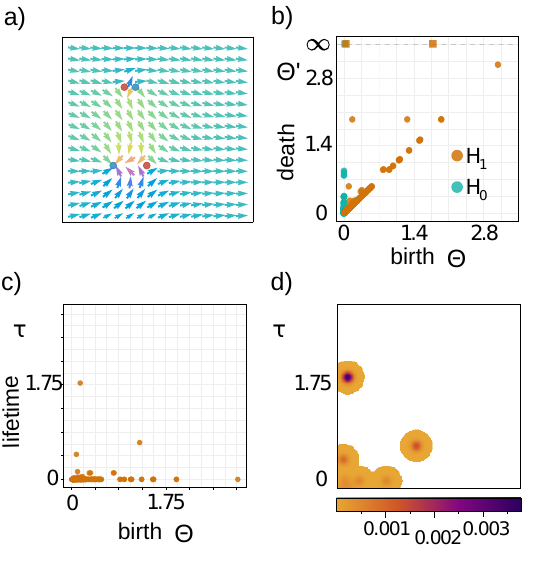}
    \caption{
    Construction of persistance diagrams and persistent images.
    a) Example of a spin configuration with two vortex pairs at 
    zero temperature (same color coding as in 
    \protect\ref{fig:vortex_pair}).
    b) Corresponding persistence diagram showing ``death'' vs.\ ''birth'' 
    of features for the zero-dimensional homology group
    ${\cal H}_0$ (teal dots) and the one-dimensional homology group
    ${\cal H}_1$ (orange dots). Every point in the graph represents the (birth-death) 
    pair $(\Theta, \Theta')$ of a feature with $\Theta$ marking the 
    filtration threshold value where the feature is first created, 
    and $\Theta'$ marking the filtration threshold where it disappears
    again.  c) Resulting persistence mapping of ${\cal H}_1$, 
    showing (birth-lifetime) pairs of features
    $(\Theta, \tau = \Theta'- \Theta)$.
    d) Resulting coarse-grained persistence image, obtained by
    smoothing the persistence mapping while giving higher
    weight to features with longer lifetimes (see Eq.\
    \ref{equ:pers_surface_weight}).
\label{fig:example_PD_PI}}
\end{figure}

The creator-destructor pairs define persistence intervals
$[f(\sigma_\alpha),f(\sigma_\beta))$, which indicate the ``lifetime''
of each topological feature.  Intuitively, a long persistence interval
corresponds to a robust, physically relevant feature, whereas short
intervals typically reflect noise or small local fluctuations. If a
creator $\sigma_\alpha$ is never paired with a destructor, the
corresponding feature persists indefinitely and is assigned the
interval $[f(\sigma_\alpha),\infty)$.  The full set of persistence
intervals constitutes the persistence barcode.  An equivalent
representation is the persistence diagram (PD), a set of points
$\big(f(\sigma_\alpha), f(\sigma_\beta)\big)$ corresponding to the
feature's  ``birth'' and ``death'' along the filtration.  For each
spin configuration, this procedure produces two persistence diagrams,
one for ${\cal H}_0$ and one for ${\cal H}_1$.  In this work, we
compute persistence diagrams for the spin configurations using the
algorithm of Edelsbrunner, Letscher, and
Zomorodian\cite{standard_algorithm, ph_roadmap}. An example of a PD
is shown in Fig.~\ref{fig:example_PD_PI}b).
 
Since PDs contain a large amount of detailed information, including many
short-lived features, we apply a weighted coarse-graining proposed by
Adams \etal\cite{Adams_PI} to obtain a compact representation.  
The PD is first mapped onto a smooth ``persistence surface'' by
replacing each point with a Gaussian centered at $(\Theta_\alpha,
\tau_\alpha) := \big(f(\sigma_\alpha), f(\sigma_\beta) -
f(\sigma_\alpha)\big)$, where $\Theta_\alpha$ is the birth scale and
$\tau_\alpha$ the lifetime of the feature. The resulting surface
is defined as
\begin{equation}
    \label{equ:pers_surface}
    \rho_{\text{PD}}(\Theta,\tau) 
       =   \frac{1}{2 \pi \sigma^2} \sum_{\alpha \in \text{PD}}
           \text{e}^{\big( (\Theta - \Theta_\alpha)^2 
           + (\tau - \tau_\alpha)^2 \big)/2 \pi \sigma^2} w(\Theta,\tau),
\end{equation}
with Gaussian width $\sigma=0.1$ and the weighting function\cite{Adams_PI}
\begin{equation}
    \label{equ:pers_surface_weight}
    w(\Theta,\tau) = \left\{ 
    \begin{array}{ll}\tau/\tau_\text{ref} & \text{for} \: 
    \tau < \tau_\text{ref} \\ 1 & \text{for} \: \tau \ge \tau_\text{ref} \end{array},
    \right.
\end{equation}
which is chosen to suppress features with short lifetimes, reflecting the 
intuition that such features are less relevant. The parameter
$\tau_\text{ref}$ is set to the maximum lifetime observed across all 
spin configuration,
$\tau_\text{ref}=\max(f(\sigma_\beta) - f(\sigma_\alpha))$, 
ensuring a consistent
comparison across configurations.  Finally, $\rho_{\text{PD}}(\Theta,\tau)$ 
is discretized on a regular grid of pixel size $a=1/L$, yielding the
persistence image (PI), which provides a compact and robust
representation of the topological structure and serves as a basis for
comparing different spin configurations.
While we use own code for most of this analysis, the final step 
of mapping the point sets of the PDs to the PI is done with the
`Persim' package from the scikit-TDA library\cite{scikitTDA_persim}, 
applying the customized weighting function introduced in Eq.\
\ref{equ:pers_surface_weight}.

In addition to computing persistence diagrams, we also directly analyze
the cubical complexes ${\cal S}_\Theta$ themselves by studying how
various characteristic quantities evolve as a function of the
filtration parameter $\Theta$. This complementary analysis provides
a more direct view of the geometrical and graph-theoretical structure
of the complexes at different stages of the filtration.

Specifically, we  examine the full complex ${\cal S}_\Theta$ as well
as two important subsets: The set ${\cal C}_\Theta$ of all plaquettes
in ${\cal S}_\Theta$, which contains only elements with nonzero
measure, and the set ${\cal E}_\Theta$ of all edges in ${\cal
S}_\Theta$, which has the structure of an undirected graph.

For the full complex ${\cal S}_\Theta$, we evaluate the Euler-Poincar\'e
characteristic, $\chi_{_{\cal S}}$, which is derived from the
Betti-numbers $\beta_i$ introduced earlier and can be calculated as
\begin{equation}
    \label{equ:euler_chara_def}
    \chi_{\cal S} = \beta_0 - \beta_1 = N_V - N_E + N_P,
\end{equation}
where $N_V$, $N_P$, $N_P$ are the numbers of vertices, edges, and
plaquettes in ${\cal S}_\Theta$, respectively.

For the plaquette subset ${\cal C}_\Theta$, we compute the scalar
Minkowski functionals in two
dimensions\cite{Mecke_Minkowski_functionals}, which include the total
area $A$, the total perimeter $P$, and the Euler characteristic
$\chi_{_{\cal C}}$ of the set ${\cal C}_\Theta$. The latter is in this
case defined as the number of connected components minus the number of
holes and can be calculated similar to Eq.~(\ref{equ:euler_chara_def}) as
\begin{equation}
    \chi_{\cal C} =  N_V^ {({\cal C})} - N_E^ {({\cal C})} + N_P,
    \label{equ:euler_minkowski}
\end{equation}
where $N_V^ {({\cal C})}$ and $N_E^ {({\cal C})}$ are the numbers of
vertices and edges in ${\cal S}_\Theta$ that are in direct contact
with a plaquette.

For the edge subset, i.e., the graph ${\cal E}_\Theta$, we calculate
the average diameter and radius of all connected components $G_i$ of
${\cal E}_\Theta$:
\begin{align}
  \label{equ:diameter}
   D &= \frac{1}{N_c} \sum_{i=1}^{N_c}
   \max_{v_j,v_k \in G_i} d(v_j, v_k), \\
   \label{equ:radius}
   R & = \frac{1}{N_c} \sum_{i=1}^{N_c}
   \min_{v_j \in G_i} \big( \max_{v_k \in G_i} d(v_j, v_k) \big).
\end{align}
Here $v_j, v_k$ are vertices, $d(v_j,v_k)$ their graph distance, i.e.,
the minimum length of paths connecting them, and $N_c$ denotes the
number of connected components in ${\cal E}_\Theta$. In addition, we
calculate a third Euler characteristic, 
\begin{equation}
    \chi_{\cal E} = \frac{1}{N_c} \sum_{i=1}^{N_c} \left( N_V^{G_i} - N_E^{G_i} + N_P^{G_i}\right),
    \label{equ:euler_edge}
\end{equation}
where $N_V^{G_i}$, $N_E^{G_i}$, and  $N_P^{G_i}$ denote the numbers of vertices, 
edges, and plaquettes associated with connected components $G_i$ of 
${\cal E}_\Theta \cup {\cal C}_\Theta$.

Analyzing these quantities (Eqs.\ (\ref{equ:euler_chara_def} -
\ref{equ:euler_edge}) enables an interpretable quaracterization of
both topologically long-lived features and more fine-grained
structures, including nested features. It thereby highlights relevant
length scales and structural characteristics that might otherwise be
overlooked.

\section{Machine learning approach}
\label{sec:wgan}

\begin{figure}[tb]
\centering
    \includegraphics[width=.9\columnwidth]{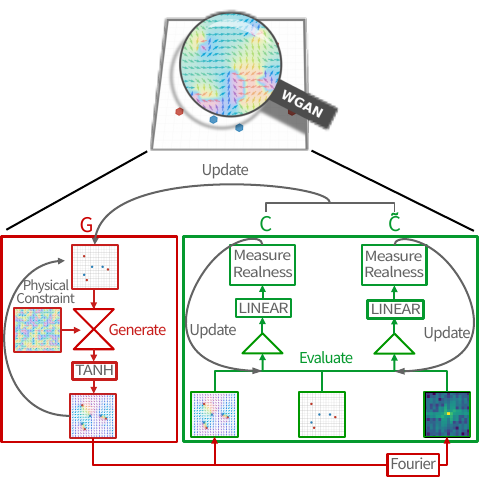}
    \caption{Illustration of the back-mapping concept 
    based on a physics-informed conditional Wasserstein
    Generative Adversarial Neural Network (WGAN). A prescribed
    defect configuration (shown here for the two dimensional XY model) is
    provided as input to the generator (see Fig.~\ref{fig:WGAN_g}), which
    produces a corresponding microscopic spin configuration. During
    training, the generated configuration is evaluated by the critics
    (Fig.~\ref{fig:WGAN_c}), which output a realism score (details see text). 
    These scores are used in a feedback loop to update the generator
    parameters.  
    \label{fig:WGAN_full}}
\end{figure}

The goal of the present work is to develop and train generative artificial neural
networks (ANNs) capable of producing representative sets of spin
configurations for given defect configurations -- encoded as winding
number distributions -- in the two-dimensional XY model, see the
schematic sketch in Fig.~\ref{fig:WGAN_full}. Here spin
configurations are represented as two-dimensional vector fields with
values in $[-1,1]$ rather than angles $\vartheta \in [0, 2 \pi]$ to
avoid difficulties in learning periodic
functions\cite{Periodic_function_Ziyin}.
The training data consist of normalized vector fields of 
the form $(\cos(\vartheta),\sin(\vartheta))$. In contrast, no
normalization constraint is enforced on the generated outputs,
which may be rescaled a posteriori if necessary.

While generative models are well-established, constructing such models
using artificial neural networks is nontrivial. Standard ANNs are
naturally designed as deterministic function approximators, whereas
generative ANNs must learn an entire probability distribution from a
finite set of samples. This requires enforcing pro\-ba\-bi\-li\-ty
normalization and, in many cases, estimating sample likelihoods in
high-dimensional spaces.  As a result, the architectures and training
processes of neural-network-based generative models are considerably
more complex than those of conventional ANN applications.

In this section, we present the general structure of the neural
networks used in this work. All models are implemented and trained
using 'PyTorch v2.2.2'~\cite{NEURIPS2019_9015}.  We first introduce
and explain the architecture and then describe the corresponding
training procedure.

\subsection{Network Architecture} 

\label{subsec:wgan_arch}

As the framework for our ANN approach, we use a conditional
Wasserstein Generative Adversarial Network (cWGAN)~\cite{WGAN},
which extends the original Generative Adversarial Network (GAN)
framework\cite{NIPS2014_5ca3e9b1}.  WGANs are known to exhibit more
stable training behavior than simple GANs and to be less prone to so-called ``mode collapse''\cite{mode_collapse}.  Although WGANs typically require relatively long training times, once trained they can generate
high-quality samples in a single forward pass -- different from
more recent generative approaches such as diffusion models, which
typically require many iterative steps to generate a single
configuration\cite{GanVSDiffusion}. Therefore, WGANs are well
suited for time efficient simulation and data analysis support.

A WGAN, as illustrated in Fig.~\ref{fig:WGAN_full} consists of two
types of neural networks: a generator $G$, which  produces synthetic
data from a given input, and a critic $C$, which is trained to
distinguish generated (artificial) data from real data. In the present
work, we extend this standard setup by introducing two critic
networks, $C$ and $\tilde{C}$, each specialized to assess different
aspects of the generated configurations as explained below.
During training, the generator $G$ iteratively improves the realism of
the generated data based on the feedback from both critics, while each
critic simultaneously learns to better discriminate between real and
generated samples.  Ideally, the trained network $G$ produces
configurations that are statistically indistinguishable from real
data. In practice, however, the adversarial setup of the networks makes GAN
training highly unstable.  Although the WGAN formulation alleviates
some of these issues, the training dynamics remain very sensitive to
the choice of hyperparameters~\cite{NEURIPS2019_9015,WGAN}.

\begin{figure}[tb]
    \centering
    \includegraphics[width=1.\columnwidth]{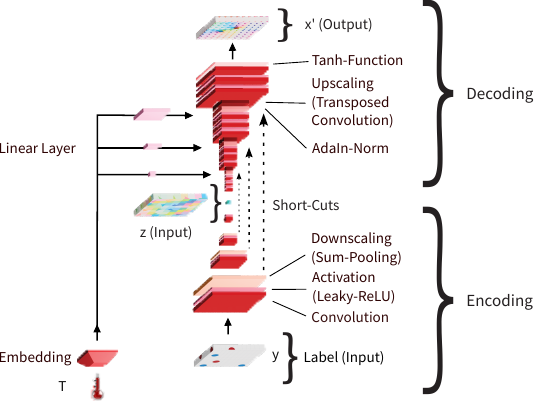}
    \caption{
    Architecture of the generator network $G(z|y,T)=x'$ based
    on a U-net architecture. The encoder maps the winding-number
    distribution $y$ and the embedded temperature labels $T$ to
    to a high-dimensional latent representation at the bottleneck.
    Gaussian noise $z$ is injected at this stage, and the combined
    latent features are subsequently decoded to generate the output
    spin configuration $x'$. Short cut connections transfer information
    from the encoder to the corresponding decoder layers at
    each resolution.}
    \label{fig:WGAN_g}
\end{figure}

The generator takes as input a two-dimensional winding number distribution $y$ (encoded as an image in the same resolution as the output) and an embedded temperature label $T$, and combines this with 
low-resolution noise $z$ to produce a generated data sample $x'=G(z|y,T)$.
Specifically, our generator network (Fig.~\ref{fig:WGAN_g}) is based 
on a U-net  architecture\cite{Unet} with an encoder and a decoder
connected by a bottleneck, and additional skip (``shortcut'')
connections between corresponding encoder and decoder layers
at each resolution. Gaussian-distributed noise $z$ is injected in the bottleneck region to generate stochastic
variability while avoiding repeated noise re-injection through the 
skip connections. The U-net captures both local spin 
correlations and global vortex structures thanks to its exponentially 
growing receptive field, i.e., region of influence in the input. 
Skip connections help to preserve fine-grained features of spin 
configurations. This multiscale  design is flexible with respect 
to lattice size  and defect number.

The winding-number input $y$ is directly encoded by the 
U-net, while the discrete temperature information is first mapped to a
low-dimensional continuous representation via an embedding layer
and later injected in the decoder via adaptive instance 
normalization(AdaIN)\cite{AdaIn}, following the style-modulation 
strategy introduced in StyleGAN\cite{StyleGAN}. The encoder captures 
the defect structure at multiple length scales and transfers this 
information to the decoder through the skip connections.
It consists of multiple blocks with progressively decreasing
spatial resolution and increasing feature dimension, up to a depth
$D_G$. Each block contains a convolutional layer followed by leaky rectified linear unit (LeakyReLU) activations\cite{Nair_relu, 
Maas_leaky_relu}  and summation-pooling for dimensional reduction, 
with the block input added to the output. LeakyReLU was chosen because 
it empirically improves convergence and stability.

The decoder upsamples the latent information in the bottleneck region 
including the injected noise, through $D_G$ layers. Each layer consists 
of an upscaling operation followed by three convolutional layers with
LeakyReLU activations. Interpolation-based upsampling was tested to 
avoid aliasing artifacts associated with transposed
convolutions\cite{checkerboardDeconv}, but proved unsuitable for
capturing the discontinuous nature of the vector field and its
underlying topology, disrupting the learning process.  The embedded
temperature is injected at each decoder layer via AdaIN, which
normalizes the feature statistics with a regularization hyperparameter
$\epsilon$. A final tanh activation constrains the output $x'$ to 
$[-1,1]$, enabling the generation of two-dimensional 
normalized spin vector fields $\hS(\theta)$.

The generator is trained by minimizing its loss function 
via backpropagation, updating the parameters of all layers
accordingly. The generator loss $\mathcal{L}_{G}$ is 
chosen as
\begin{equation}
    \label{equ:Loss_G}
    \begin{split}
    \mathcal{L}_{G} := &
     -\frac{1}{2}(\mathbb{E}[f_C(x'|y,T)] 
     + \mathbb{E}[f_{\tilde{C}}(x'|y,T)])\\
    & + \beta \: \mathbb{E} \Big[\textstyle 
      \sum_{r=1}^N \alpha_r\vert \tilde{k}_r - 
      k_r^{\text{target}}
      \vert^2 \Big],
    \end{split}
\end{equation}
where the first two terms are standard WGAN contributions, driving the
generator to maximize the output of the two critics and thereby
produce samples that are indistinguishable from real data (the
expectations are taken over the generated samples). The third term is
a physics-informed $L_2$ loss with regularizing factor $\beta$
that penalizes deviations between the target (input) winding number
distribution $\{k_r^{\text{target}}\}$ and the winding number
distribution, $\{k_r\}$, of the generated spin configurations. Since
the exact winding-number definition, Eq.~(\ref{equ:vortex_discrete})
is non-differentiable and therefore incompatible with backpropagation,
we approximate $k_r$ by the differentiable expression

\begin{equation}
    \label{equ:cross_prod}
    k_r \approx \tilde{k}_r 
      = \frac{1}{4} \sum_{\{i,j\} \in \Box_r} 
      \left(S_{i_x}S_{j_y} - S_{j_x}S_{i_y}\right),
\end{equation}
which is based on the cross products of plaquette spins $S_{i,j}$.
In simulations, the mean deviation $\Delta k =
\langle|\tilde{k}_r- k_r| \rangle$ between $\tilde{k}_r$ and the true
winding number, shown in Fig.\ 1 in SM, is mostly of order $3 \times
10^{-3}$ and does not exceed 1 \%.  Nevertheless, the approximation 
(\ref{equ:cross_prod}) implies that the local loss terms
$\vert \tilde{k}_r - k_r^{\text{target}} \vert^2$  do not vanish
exactly, even for configurations with the correct defect structure.
Moreover, most plaquettes do not host defects but still contribute to
the loss.  To mitigate their influence, we introduce weighting factors
$\alpha_r$: plaquettes corresponding to true or predicted defect
locations are weighted by a factor $N$, while all other plaquettes are
assigned unit weight.
 
 \begin{figure}[tb]
    \centering
    \includegraphics[width=01.\columnwidth]{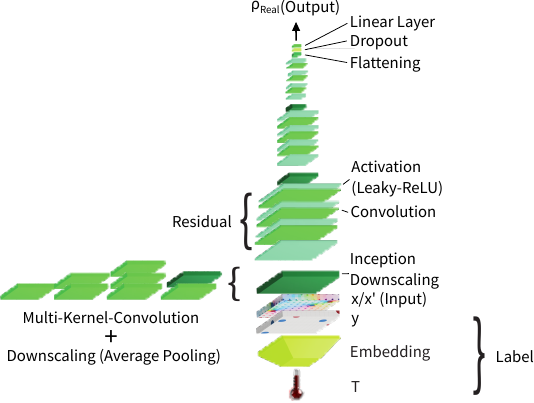}
    \caption{Architecture of the critic network $C(x,x'|y,T)$. It
    takes either the generated output $x'$ or training data $x$ as 
    input concatenated with the winding number label $y$ and 
    temperature label $T$.
    \label{fig:WGAN_c}}
\end{figure}

The second component of our WGAN consists of two critic networks, $C$
and $\tilde{C}$, which evaluate the generator output. Unlike standard
GANs\cite{GAN} that classify inputs as ``real'' or ``fake'', a WGAN
critic measures the difference between the distributions of generated
data $x'=G(z|y,T)$ and real samples $x$ for given generator input
$\{y,T\}$.  It is based in the Wasserstein metric, a concept borrowed
from optimal transportation theory\cite{Wass, WGAN}, which provides
a distance measure between distribution functions $P_i$ and $P_j$,
similar to the Kullback-Leibler distance\cite{KulLeib}. The
1-Wasserstein distance can be calculated as\cite{Villani_2003}
\begin{equation}
\label{equ:Wass}
    W(P_i,P_j) = 
       \text{sup}_{f} \big[\mathbb{E}_{z \sim P_i}[f(z)] 
       - \mathbb{E}_{z \sim P_j}[f(z)] \big],
\end{equation}
where the supremum is over all 1-Lipschitz functions $f$. Each 
critic is trained to approximate the function $f$ that maximizes this
distance. The loss for each critic $\hat{C}$
(with $\hat{C}=C$ or $\tilde{C}$) is
\begin{equation}
        \label{equ:Loss_C}
    \mathcal{L}_{\hat{C}} 
       := -(\mathbb{E}[f_{\hat{C}}(\hat{x}'|y,T)]
         - \mathbb{E}[f_{\hat{C}}(\hat{x}|y,T)]) + GP,
\end{equation}
where GP is a gradient penalty term\cite{impWGAN}
\begin{equation}
    \label{equ:GP}
    GP :=  \lambda\cdot\underset{\overline{x}
      \sim \mathbb{P}_{\overline{x}}}{\mathbb{E}}
       [(\vert\vert\nabla_{\overline{x}}
          f_C(\overline{x}|y,T) \vert\vert_2-1)^2]
\end{equation}
enforcing the 1-Lipschitz constraint and, as a beneficial side effect,
also aids in stabilizing training. Here $\overline{x}$ is sampled along 
linear interpolations between real and generated 
data\cite{impWGAN}, $\{x,x'\};\{\tilde{x},\tilde{x}'\}$,
and $\lambda$ is a hyperparameter. In contrast to discriminator
components of standard GANs, which assign discrete scores $\{0,1\}$ to
its input, the critic in a WGAN thus outputs a continuous scoring
function, which is trained to be largest for real data (see also
Eq.~(\ref{equ:Loss_G})) and to maximize the difference in outputs between
real and fake data. This improves the stability of the training and
was shown to help preventing  mode collapse\cite{modecol}.

\begin{figure}[tb]
    \centering
    \includegraphics[width=0.9\columnwidth]{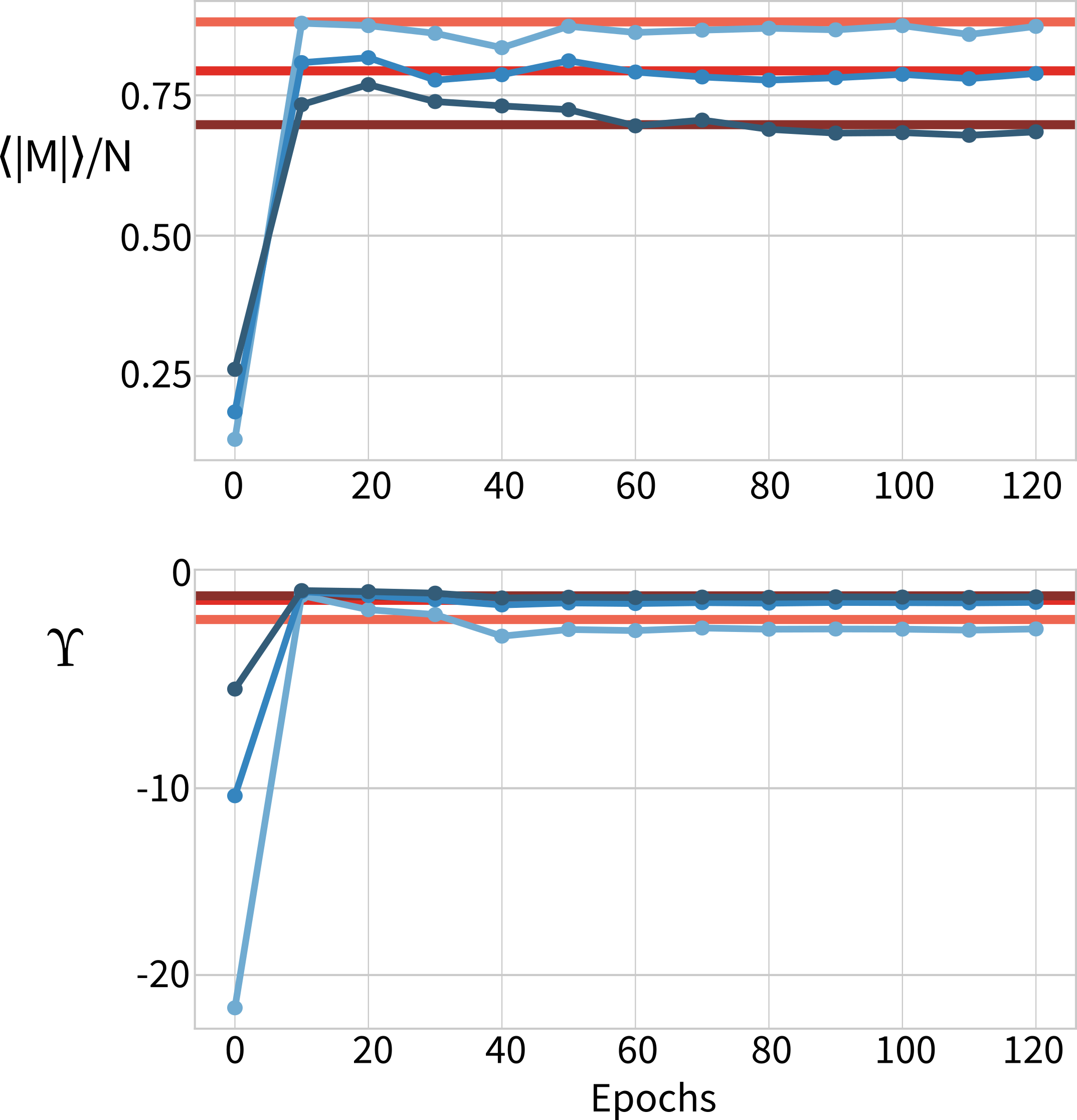}
    \vspace{-\baselineskip}
    \caption{Training progress of generator (blue) compared to the
    simulated baseline (red) for the intrinsic mean magnetization per
    spin $\langle|M|/N\rangle$ and the helicity modulus $\Upsilon$ at temperatures \mbox{$T=0.2 \, J/k_B$} (light), $T=0.5 \; J/k_B$ (medium) and $ T=0.8 \, J/k_B$
    (dark). 
    \label{fig:epoch_train}}
\end{figure}

We employ two critics with identical architectures
(Fig.~\ref{fig:WGAN_c}), one focusing on real space
characteristics of the spin configurations and one focusing 
on Fourier space. The real-space based critic $C$ receives the 
two spin vector fields of the generated $x'$ and the training spin 
data $x$, while the Fourier-space critic $\tilde{C}$ receives the 
normalized logarithmic Fourier power spectra
$\tilde{x};\tilde{x}' =
2\left(\frac{\log_{10}
|F(x;x')|^2}{\max\left(\log_{10}|F(x;x')|^2\right)}-0.5\right)$.
The Fourier-space input enhances spectral resolution, thereby resolving
defect-induced spin variations more accurately, and improves enforcement of
the winding-number distributions. 

Both inputs are concatenated with the generator labels $y$ and $T$.
Each critic begins with an inception layer\cite{inception}, followed
by LeakyReLU activations, to extract
multiscale features while reducing dimensionality. Intended to capture
different correlation lengths, the inception layer uses convolutions
with kernel sizes 1–5 and average pooling for coarse information
mapping. Each block concludes with a residual module\cite{ResNet} 
composed of three convolutional layers and by LeakyReLU activations
for additional feature selection. Finally, the hidden features are 
flattened and mapped through a linear layer to produce a continuous, 
scalar realism score for each input.

\begin{figure}[tb]
     \centering
    \includegraphics[width=\linewidth]{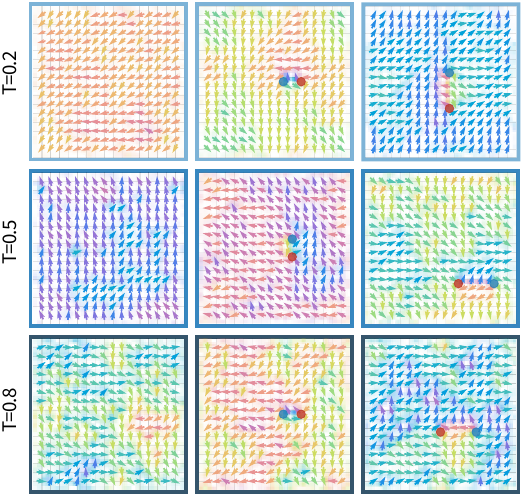}
\caption{Examples of generated spin configurations for \mbox{$T=0.2 J/k_B$}
   (top), \mbox{$T=0.5 J/k_B$} (middle), \mbox{$T=0.8 J/k_B$} (bottom)
   without defects (left), with two close defects (middle),
   and two distant defects (right) (same color coding as in
   Figure \protect\ref{fig:vortex_pair}).
   \label{fig:generated_spin_lattices_main}
    }
    \vspace{-0.3cm}
\end{figure}

\begin{figure*}[t]
\centering
  \hspace{-1.0cm}
  \includegraphics[width=\linewidth]{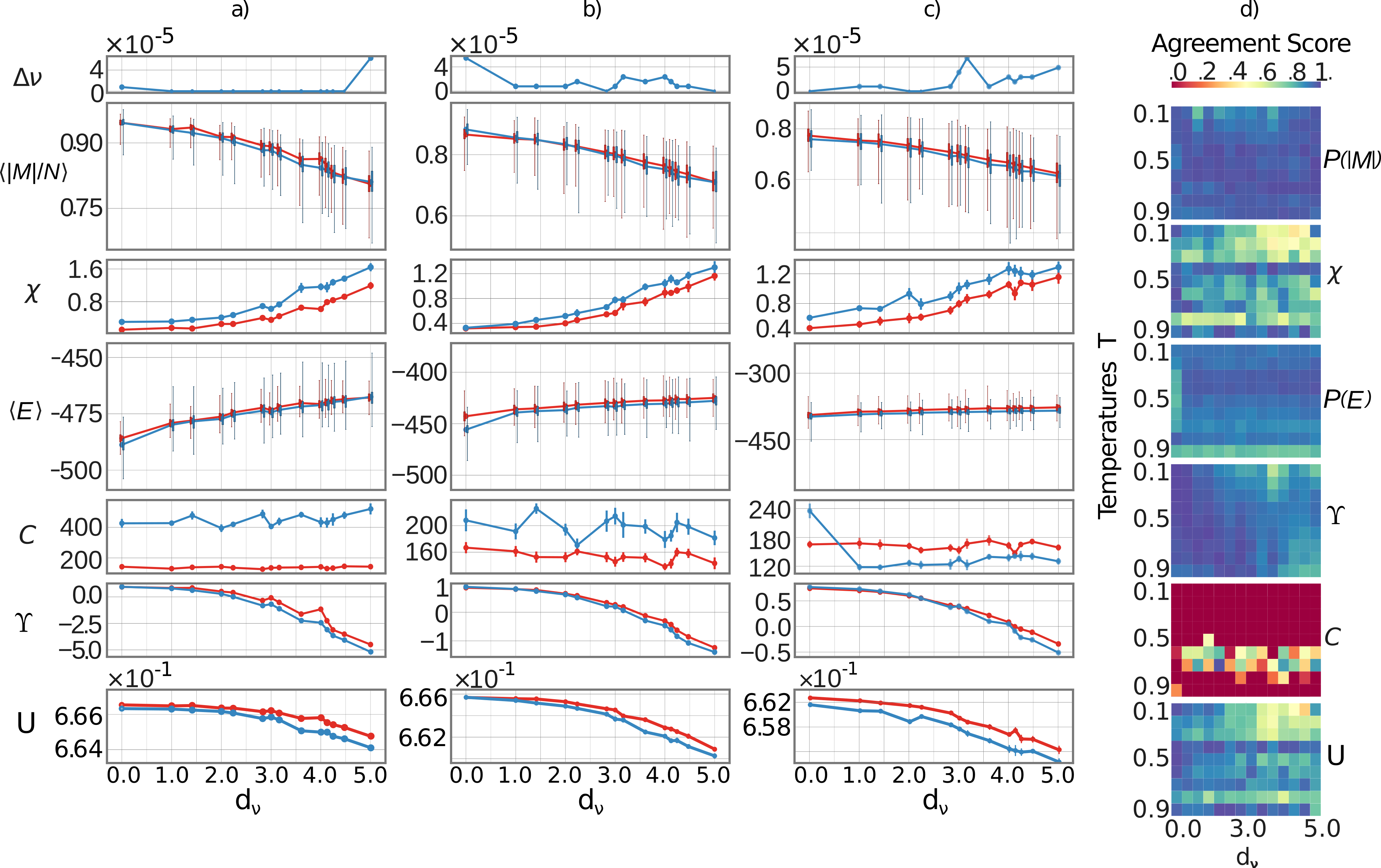}
 \vspace*{-1\baselineskip}
  \caption{ Physical observables versus 
  defect pair distance $d_{\nu}$ for a) $T=0.2 \, J/k_B$, 
  b) $T=0.5 \, J/k_B$, and c) $T=0.8 \, J/k_B$,
  based on the analysis of 1000 simulated (red) and 
  generated (blue) spin lattice configurations per parameter
  combination.
  The observables are:
  defect configuration difference $\Delta\nu$ 
     (Eq. (\protect\ref{eq:winding_error})),
  mean magnetization per spin $\langle|M|/N\rangle$,
  magnetic susceptibility $\chi$ 
  (Eq.~(\protect\ref{equ:magnetic_susceptibility})), 
  mean energy $\langle E \rangle$,
  specific heat $C$  (Eq.~(\protect\ref{equ:specific_heat})), helicity modulus $\Upsilon$ 
    (Eq.~(\protect\ref{equ:helicity_modulus})),
  and the Binder cumulant $U$ 
    (Eq.~(\protect\ref{equ:binder}))
  The vertical bars in the graphs for $\langle |M|/N \rangle$
  and $\langle E \rangle$ indicate the range of the respective 
  distributions omitting outliers.
  d): Heat map of corresponding agreement score (full
  temperature range). 
  }
\label{fig:main_physical_analysis}
\end{figure*}

\subsection{Training process and hyperparameter tuning}
\label{subsec:training}

For the results discussed in the next section, networks were trained
for 120 epochs in batches of 64 on an 'NVIDIA RTX 3080', using a
dataset of $315 000$ spin lattices. To ensure a well-balanced and informative
training set, each defect configuration was paired with an equal
number of spin configurations for every temperature label. In
addition, approximately one third of the dataset consists of
defect-free configurations, which enhances the model’s ability to
accurately capture thermal fluctuations.\cite{footnote1}
The Fourier-space inputs $\tilde{x},\: \tilde{x}'$ were separated into
two channels, one for each spatial direction, to capture
direction-specific gradients.  Network weights were initialized using
PyTorch defaults and trained with the ADAM optimizer\cite{ADAM} with
parameters $\beta_1 = 0.9$ and $\beta_2 = 0.999$ at a learning rate
$10^{-4}$. To prevent overtraining on the winding number loss, this
term was introduced after 40 epochs with $\beta = 100$.
Critics were trained with $N_{C;\tilde{C}/G}= 10$
critic updates per generator update. This asymmetry in the training
protocol ensures meaningful gradients.

While the training was mostly stable for this choice of
hyperparameters, the quality of the generated spin configurations --
particularly correlation functions  at low temperatures and defect
positioning at higher temperatures -- was sensitive to the precise
choice of parameters in the loss functions. Good results were obtained
using $\alpha_i = 100$ for the defect loss parameter in the generator
(Eq.~\ref{equ:Loss_G}), $\lambda = 1$ for the critic gradient
penalty (Eq.~\ref{equ:Loss_C}), and $\epsilon =0.5$ as the AdaIN
regularization factor (Fig.\ \ref{fig:WGAN_g}).

For cWGANs, the generator loss (Eq.~\ref{equ:Loss_G}) does not
directly reflect the physical realism of the generated data, as it
only measures performance relative to the critics. Therefore, training
progress is monitored using the magnetization $\langle|M|\rangle$
and the spin-stiffness or helicity modulus $\Upsilon$
as indication of the overall alignment of the spins. As shown in the
example of Fig.~\ref{fig:epoch_train}, both quantities improve rapidly
in early epochs, with further refinement occurring after the
defect-loss term is introduced at epoch 40.

\section{Results and Network Validation}

\label{sec:validation}

In this section we evaluate the ability of our neural network
to learn and reproduce physical and topological features based on the
measures introduced in the sections {\em Thermodynamic Properties} and
{\em Topological Data Analysis}.  Specifically, we consider
homogeneous batches of defect-pair configurations with fixed
defect-pair distances $d_{\nu}$ and temperatures $T$.  Some
representative examples are shown in
Fig.~\ref{fig:generated_spin_lattices_main}.

We first verify that the topological constraints are faithfully
implemented in the generated configurations. The top panels in
Fig.\ \ref{fig:main_physical_analysis} a)-c) 
show the overall deviation 
between the target winding number $k_r^{\text{target}}$
and the achieved winding number $k_r$,
quantified by
\begin{equation}
\label{eq:winding_error}
    \Delta\nu = \frac{1}{N}\sum_{r}^N|k_r-
     k_r ^{\text{target}} |
\end{equation}
for different defect distances and temperatures. In all cases, 
the defect distribution error
$\Delta \nu$ remains of order $\Delta \nu \sim {\cal O}(10^{-5})$, 
demonstrating that the generator successfully enforces
the constraints imposed by the physical loss function, 
Eq.~(\ref{equ:Loss_G}).  

Next we turn to the statistical analysis of physical observables
defined in Section {\em Thermodynamic Properties}.
Fig.~\ref{fig:main_physical_analysis} compares statistical measures
obtained from generated data (blue) and simulated data (red) for
defect-pair distances $d_\nu$ ranging from 0 to 5 and  three
temperatures $T=0.2 \, J/k_B$, $T=0.5 \, J/k_B$, and $T=0.8 \; J/k_B$,
all below the Kosterlitz-Thouless transition ($T_{KT} = 0.8929(1) \;
J/k_B$)\cite{Hasenbusch2005XY}).  To quantitatively assess 
agreement between the generated and simulated ensembles, we compute an
``agreement score'' for the distributions of energy and magnetization,
$P(E)$ and $P(|M|)$, as well as for the susceptibility $\chi$, the
specific heat $C$, the spin stiffness $\Upsilon$, and the Binder
cumulant $U$, as described in
the appendix.  Agreement for the distributions ($P(E)$ and $P(|M|)$)
is quantified using the the mean-normalized Wasserstein-1 distance
(see Appendix, Eq.~(\ref{equ:acc_W1})), while agreement for the other
other observables is based on the normalized root mean square error
$\text{nRMSE}$ (see Appendix, Eq.~(\ref{equ:acc_observables})).  The resulting scores are shown as
heatmaps in Fig.~\ref{fig:main_physical_analysis} d). 

The generated spin configurations show good agreement for the mean
magnetization per spin $\langle|\mathbf{M}|/N \rangle$, the magnetic
susceptibility $\chi$, and the Binder cumulant $U$, with
agreement scores of  $0.785-0.994$,
$0.42-0.99$, and $0.849-0.957$, respectively. This demonstrates
the network's ability to reconstruct the intrinsic,
temperature-dependent alignment of spin fields including fluctuations.
Similarly, the
agreement scores for the energy distribution, $P(E)$, are consistently
high (0.778-- 0.984), indicating that the generated distributions are
centered at the correct energy values $E$. In contrast, the specific
heat $C$, which is directly linked to the variance of the energy,
shows significant deviations from the simulation baseline. Despite
extensive efforts, this discrepancy could not be fully eliminated.
Introducing an additional Fourier-based critic into the GAN
architecture (see the previous section) reduced deviations by
enforcing longer-wavelength spin waves and suppressing spurious local
fluctuations\cite{spectraldist_gan_problem, spectraldist_gan_improve},
but further improvement was not achieved. A comparison of
selected physical observables computed from configurations generated
with and without the Fourier critic is provided in the Supplementary
Material, Figs.~4 and 5.

To further investigate this discrepancy, we examined the
distributions of energy $E$ and magnetization per spin $|M|/N$
(Supplementary Material, Fig.\ 3).  While the generated magnetization
distributions largely match the simulated ones, with occasional
deviations possibly reflecting stochastic training instabilities, the
generated energy distributions tend to be too broad and feature
temperature-dependent fat tails.  This behavior may reflect the
composite nature of energy fluctuations in the XY model, which arise
from a combination of extended spin-wave modes, transient localized
fluctuations, and stable vortex excitations, each with distinct
temperature dependence.  Accurately reproducing the energy variance
therefore requires the network to capture multiple contributions
simultaneously.  By contrast, magnetization fluctuations are dominated
by pairwise spin correlations, which are comparatively easier to
learn.

\begin{figure}[t]
\centering
    \includegraphics[width=1.\linewidth]{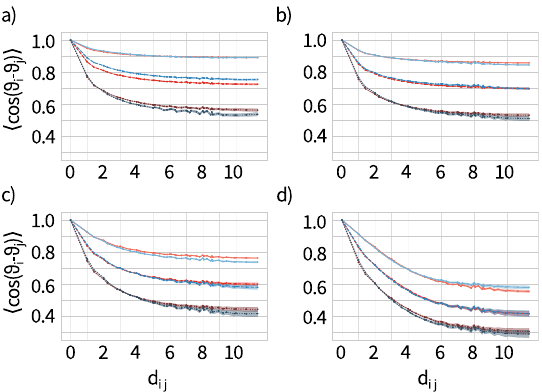}
    \caption{Spin-spin correlation function 
    $\langle \hS_i \hS_j \rangle = \cos(\vartheta_i -
    \vartheta_j) \rangle$ as a function of spatial distance 
    $d_{ij}$ of the spins as obtained from generated (blue) and 
    real simulated (red) spin configurations for $T=0.2 \, J/k_B$ 
    (top light curves), $T=0.5 \, J/k_B$  (middle curves)
    and $T=0.8 \, J/k_B$ (bottom dark curves) and different defect
    configurations:
    (a) no defects $d_{\nu}=0$, 
    (b) two close defects $d_{\nu}=1$
    (c) two defects at medium distance $d_{\nu}=3$, 
    (d) two distant defects, $d_{\nu}= 5$.
    Shadings indicate statistical errors.
    \label{fig:main_correlation_function}
    }
\end{figure}

Despite these discrepancies in energy fluctuations, other
observables that probe collective and long-range behavior are 
reproduced with high fidelity. In particular,
the helicity modulus $\Upsilon$, which characterizes the
robustness of the spin lattices against twisting boundaries, is
reproduced with high accuracy, yielding agreement scores in the range
$0.668-0.999$, even in regimes where $\Upsilon$ becomes negative
and the spin system is unstable. 
Furthermore, spin-spin correlations 
\mbox{$\langle \hS_i \cdot \hS_j \rangle 
= \langle \cos(\theta_i - \theta_j) \rangle$}
are well captured by the generated configurations, as shown in
Fig.~\ref{fig:main_correlation_function}
(see also Fig.\ 1 in the Supplementary Material (SM) for additional data).
Minor deviations at larger defect distances mainly reflect the
deviations of the baseline quantity, 
$\langle \hS \rangle^2 \sim \langle |M|/N \rangle^2$.

Overall, these results demonstrate that the generative network 
successfully reproduces physically realistic ensembles of spin
configurations containing defect pairs with different distances
over a broad range of temperatures, from low to near-critical regimes.

This indicates that the network has learned a temperature-conditioned
representation of constrained spin ensembles with fixed defect positions,
as further supported by the Pearson correlation coefficient matrix of
the mean and variance of the embedded temperature labels
(Supplementary Material, Fig. 2).

\begin{figure}[t]
\centering
\includegraphics[width=1.\linewidth]{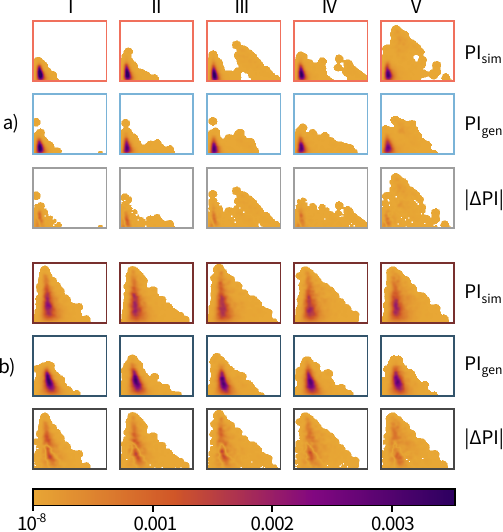}
  \caption{Mean persistence images of ${\cal H}_1$ for simulated test data
  (left), network-generated data (middle), and 
  $L_1$ difference between them (right), 
  for $d_{\nu} = 0.0, 1.0,  2 \sqrt{2}, 3.0, 5.0$ in I-V, 
  respectively. Results are shown for temperatures $T=0.2 \, J/k_B$
  and $T=0.8 \, J/k_B$ in a) and b) respectively. Coloring begins at
  the threshold value of $10^{-8}$.
  \label{fig:PI_main}
  }
\end{figure}

\begin{figure*}
\centering

\includegraphics[width=\linewidth]{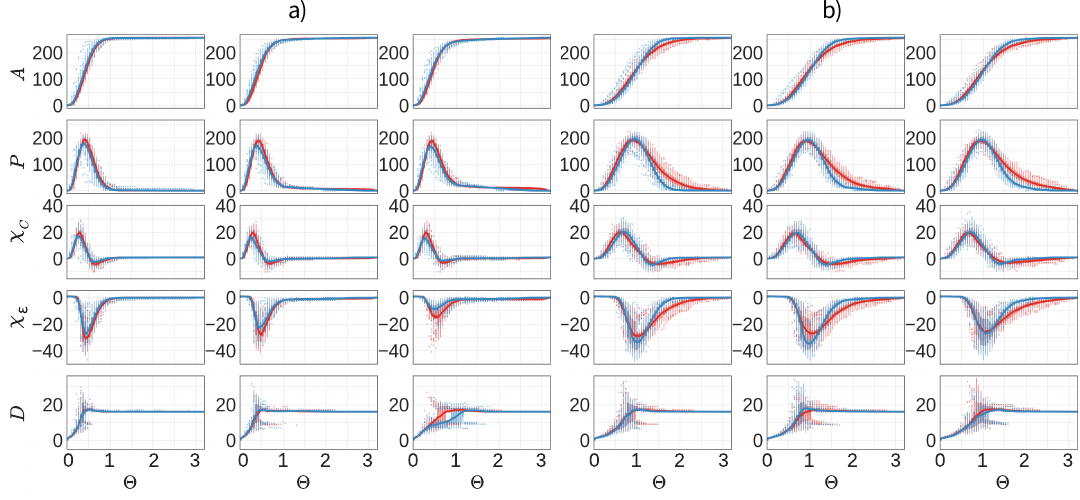}
   \caption{Geometrical, topological, and graph-based descriptors
   calculated from generated (blue) and simulated (red) data:
   Two-dimensional Minkowski measures $A$ (area), $P$ (perimeter),
   $\chi_{\cal C}$ (Euler characteristics, 
   Eq.~(\protect\ref{equ:euler_minkowski})), 
   Edge-graph based Euler
   characteristics $\chi_{\cal E}$ 
      (Eq.~(\protect\ref{equ:euler_edge})), 
   and graph diameter $D$ (Eq.~\protect\ref{equ:diameter})) 
   -- versus filtration parameter
   $\Theta$ for systems without defects (left), close defects with
   distance $d_\nu = 1$ (middle), 
   and distant defects with distance $d_\nu = 3$ (right) 
   at temperatures $T=0.2 \, J/k_B$ (a) and $T=0.8\, J/k_B$ (b). 
   Data were collected from a set of 100 spin configurations each. 
  \label{fig:TDA_main}
   }
\end{figure*}

The preceding analysis relies on thermal observables, which
provide statistical information only at the ensemble level. To assess
whether individual generated configurations are physically and
topologically realistic, we next analyze topological features using
persistent homology, as introduced in Section {\em Topological Data
Analysis}. Using the filtration function defined in
Eq.~(\ref{equ:filtration}), we construct a sequence of cubical
complexes for each individual spin lattice by varying the filtration
threshold $\Theta$,
based on the metric of minimal spin-angle differences. From these
sequences, we compute persistence images (see Fig.~\ref{fig:PI_main}),
which encode the lifetimes of homology groups as they appear and
disappear along the filtration. Details of the procedure can be found
in Section {\em Topological Data Analysis} and
Refs.\inlinecite{persistent_homology,ph_roadmap}.
Persistence images are particularly useful because they provide
compact, statistically meaningful representations of topological
behavior across batches of configurations. Fig.~\ref{fig:PI_main}
compares persistence images obtained from simulated and generated
data, $\PI_\text{sim}$ and $\PI_\text{gen}$, as well as their L$_1$
difference $|\Delta\PI| = |\PI_\text{sim}-\PI_\text{gen}|$, for
temperatures $T=0.2 \, J/k_B$ and $T=0.8 \, J/k_B$.

In the absence of defects (case i), the persistence \mbox{images} reveal 
an accumulation of early-birth ${\cal H}_1$ homo\-logy groups --
representative of spin-wave patterns -- with low persistence at low
temperature (Fig.\ \ref{fig:PI_main} left) up to $T= 0.6 \; J/k_B$
(data not shown).  As the temperature approaches criticality, these
features extend to higher persistence values, indicating complex
fractal-like correlation structures. While the network produces
similar patterns at low temperatures, it underestimates the range of
lifetimes of low-birth features in the near-critical regime,
indicating that our network does not fully capture the global 
multiscale nature of critical fluctuations. It is worth noting
that the real spin configurations exhibit such critical fluctuations
despite the (enforced) absence of topological defects.

When defects are introduced, the persistence images change
qualitatively. Depending on the distance $d_\nu$ between
defect pairs and their relative positioning, the lower-left triangle
of the persistence diagram becomes increasingly populated,
reflecting the formation of persistent non-boundary cycles
at higher thresholds $\Theta$, as one might expect in the presence of
topological vortices. This behavior is observed in both simulated 
and generated data; however, the generated configurations tend
to underestimate the persistence of these cycles. 

Overall, the persistent homology analysis indicates that the
${\cal H}_1$ features in the generated configurations decay too rapidly
during the filtration process, implying that non-boundary cycles
are filled prematurely. This limitation points to deficiencies
in capturing global spin-correlation properties that are not readily
apparent from two-point correlations alone.

To gain further insights, we extend the analysis to \mbox{additional}
topological and geometric measures derived from the cubical complex
sequences.  Results are shown in Fig.~\ref{fig:TDA_main}
for selected defect-pair distances $d_{\nu}$ and temperatures $T
\in [0.2,0.8] J/k_B $.  Agreement scores are again computed
using the Wasserstein-1 distance (Eq.~(\ref{equ:acc_W1})), evaluated for batches of 100 configurations
and averaged over the full range of $\Theta$.
 
We begin with the plaquette subsets ${\cal C}_{\Theta}$ of the
cubical complexes ${\cal S}_\Theta$, and examine the corresponding
Minkowski functionals area $A$, perimeter $P$, and
Euler-characteristics $\chi_{\cal C}$ (Eq.~\ref{equ:euler_minkowski}).
These measures capture the global geometry of spin clusters based on
their mesoscale spin alignment. As $\Theta$ increases, the system
transitions from a generic initial growth regime to a regime
influenced by topological defects.  The area increases monotonically,
while both perimeter and Euler characteristic exhibit pronounced
peaks, reflecting the growth of individual clusters that subsequently
merge into a single large cluster.  Negative values of $\chi_{\cal C}$
indicate local fluctuations that induce holes in the expanding
clusters. Increasing the temperature shifts these features to larger
values of $\Theta$, as shown in Fig.\ \ref{fig:TDA_main} b).

The generated data reproduce these trends with good fidelity,
achieving agreement scores of 0.705-0.970 for $P$, and 0.801-0.976 for
$\chi_{\cal C}$, with particularly strong agreement during the early
growth regime. In the near-critical regime, the peaks of $P$ and
$\chi_{\cal C}$ derived from generated configurations decay more
rapidly with increasing $\Theta$ than those derived from real
configurations, indicating again that the network cannot fully capture
critical correlations on large scales.
At low temperature, once the lattice sites merge into a single
cluster, the stability of holes is reflected in plateaus of the
perimeter and Euler characteristic. The magnitude of these plateaus
depends on the defect-pair distance $d_\nu$; for sufficiently large
distances, a single large hole decomposes into two metastable ones.
While the generated vortex field largely reproduces this behavior, it
appears to be perturbed by additional small-scale spin-angle
fluctuations, leading to earlier and more gradual decay of the
plateaus.

The final two quantities shown in Fig.~\ref{fig:TDA_main} 
characterize complementary aspects of the cubical complex
${\cal E}_\Theta$. The first is mean Euler characteristics, 
$\chi_{\cal E}$, of the connected components of 
${\cal E}_{\Theta} \cup {\cal C}_{\Theta}$
(Eq.~\ref{equ:euler_edge}). We focus on $\chi_{\cal E}$ 
rather than the Euler-Poincar\'e characteristics of the full complex 
$\chi_{\cal S}$ to avoid the otherwise dominant contribution
from isolated vertices (see Fig.~2 in SM).
The second quantity is the average diameter $D$ of the edge graph
${\cal E}_{\Theta}$, (Eq.~\ref{equ:diameter}),
which probes the stability of the global topology, including edge 
connectivity.

As the filtration parameter $\Theta$ increases, edge clusters
rapidly grow and merge, resulting in the growth of the diameter $D$
until it reaches the limit of the lattice size $L$. Concurrently,
holes emerge due to spin waves, thermal fluctuations, and vortices.
Elongated holes associated with spin waves produce deep negative
minima in $\chi_{\cal E}$.  For distant defect pairs with strong
long-range interaction fields, spin-waves are perturbed, leading
to a pronounced suppression of these minima.  At higher
temperatures, the sensibility of $D$ and $\chi_{\cal E}$ to 
defect structures diminishes, and the minima of $\chi_{\cal E}$ 
broaden and shift to larger $\Theta$.

The generated data reproduce most of this behavior with high accuracy,
reaching agreement scores of $0.732-0.99$  for $\chi_{\cal E}$ and
$0.891-0.992$ for $D$. At low temperature and $d_\nu =3$, the
generated data for $D$ exhibit a plateau in the regime $\Theta \sim
0.5-1$ which is not observed in the simulated data. This indicates
that, in generated configurations, edge clusters surrounding the two
defects are more distinctly separated from each other than in
simulated configurations and resist merging. A similar observation is
made for $d_\nu = 2 \sqrt{2}$ at low temperature (see Fig.~2 in SM).
Regarding $\chi_{\cal E}$, the simulated and generated curves agree
within the (large) statistical error, but the data nevertheless
suggest some trends: At low temperatures, the generated configurations
tend to slightly underestimate the minimum of $\chi_{\cal E}$.  At
near-critical temperatures, they overestimate the depth of the minimum
and the subsequent increase with increasing $\Theta$ is too fast. 
Thus the holes derived from generated spin configurations seem to
persist over a narrower range of length scales, consistent with the
observations from persistent images.

In summary, our generative U-net approach is able to generate
physically realistic spin fluctuations and topologically sound spin
lattice configurations constrained to prescribed defect distributions.
While the generated ensembles show overall good agreement with
simulated data, significant discrepancies are observed in the specific
heat, indicating that fourth-order spin-spin correlations associated
with energy fluctuations are not reproduced with the same accuracy as
lower-order correlations. Beyond this limitation, the detailed
analysis shows that our network reliably produces long-range spin-wave
structures except close to the critical point. With few physically
irrelevant exceptions, (isolated defects with distance 2-3 at low
temperatures), it can also capture the interaction between the vortex
field and the spin waves.  

We further explored the scalability of the approach to larger
lattice sizes and larger defect numbers.  Tests for $L = 32$ and $N_d
= 4$ demonstrate that the network can be used for larger system
sizes or defect numbers without requiring architectural modifications,
with only minor adjustment of the training settings.  Using a reduced
generator learning rate of $5 \times 10^{-5}$ and
$N_{C;\tilde{C}/G}=20$, we obtain accuracy comparable to the baseline
case, with the specific heat again showing the largest deviations.
Corresponding analyses are shown in Supplementary material,
Figs.\ 8-10. Finally, we assessed the transferability of the trained
network to higher defect numbers.  Selected statistical analyses and
representative spin configurations for $N_d=4$ are provided in the
Supplementary Material (Figs.~11 and 12), suggesting that the network
generalizes well to larger defect numbers, although this needs
to be investigated more systematically in future work.

\FloatBarrier

\section{Conclusion and Outlook}

In this work, we demonstrated the successful implementation of a
generative U-net architecture for reconstructing physically and
topologically consistent spin configurations of the two-dimensional XY
model. A detailed analysis of thermodynamic, geometric, and
topological observables shows that the network reproduces most
physical properties with high accuracy across a broad range of
temperatures and defect-pair distances, while also revealing
systematic limitations.
Despite the overall good agreement between generated and simulated
ensembles, a persistent discrepancy is observed in the specific heat,
indicating that energy fluctuations and higher-order correlations are
not fully captured. Addressing this issue was a primary motivation for
introducing the Fourier-based additional critic, which enforces
longer-wavelength spin structures and reduces spurious local
fluctuations. While this modification leads to partial improvement, it
does not eliminate the discrepancy, suggesting a fundamental
limitation that cannot be resolved within the present architectural
framework. It will therefore be of interest to investigate alternative
generative architectures and training strategies, as well as to extend
the analysis to other spin systems.  

A second important aspect of this work is the introduction of topological
data analysis as a diagnostic tool for spin configurations. This includes 
the detailed characterization based on cubical complex representations. 
Persistent homology, Minkowsky functionals and related topological 
and geometric measures reveal complex global spin-correlation structures,
particularly near critical points, that are not apparent from conventional 
two-point correlation functions alone. The cubical-complexed-based 
topological and geometric analysis provides a sensitive and physically 
interpretable probe of both defect-induced and fluctuation-driven 
structures, and proves essential for identifying subtle deficiencies 
in the generated configurations. 
More broadly, these results demonstrate that topological and geometric
measures offer a powerful and general framework for the quantitative
characterization of spin systems. Beyond their use as validation
tools in machine learning, these techniques offer a promising avenue
for  gaining new insights into multiscale correlation structures
underlying critical behavior\cite{XY_filtration}.  

Finally, the presented generative backmapping strategy establishes
a viable multiscale framework in which computationally efficient 
defect-particle dynamics can be combined with the reconstruction 
of detailed microscopic spin configurations. This opens the door 
to large-scale and long-time simulations of defect-mediated phenomena, 
while retaining access to microscopic physical information when needed. 
Beyond the XY model, the approach is naturally transferable to a 
wide range of systems with topological defects, including nematic 
and active liquid crystals, skyrmionic magnetic materials, 
and crystalline systems with dislocation networks. 
In this broader context, the proposed methodology provides a 
flexible and physically grounded pathway toward multiscale 
modeling of complex defect-driven phenomena.

\section{Supplementary Material}

The supplementary material encompasses twelve figures showing additional simulation results.

\section{Acknowledgements}

We acknowledge funding from the Emergent AI Center funded by the
'Carl-Zeiss-Stiftung', the TopDyn Research Initiative of Rhineland
Palatinate and funding by the Deutsche Forschungsgemeinschaft (DFG,
German Research Foundation) - Project numbers 233630050 and 465145163
- CRC/TRR 146 (project MGK) and SFB 1552 (project C2).  KES acknowledges funding from the DFG - Project numbers 405553726 and 278162697 - CRC/TRR 270 (project B12) and SFB 1242 (project B10).

\section{Author Declarations}

\subsection{Conflict of Interest}

The authors declare no competing interests.

\subsection{Author Contributions}

{\bf Kyra Klos:}
Methodology (equal); Software (equal); Formal Analysis (lead); Investigation (lead); Data Curation (lead); Writing - Original Draft (lead); Writing - Review \& Editing (supporting).
{\bf Jan Disselhoff:}
Methodology (equal); Software (equal); Formal Analysis (supporting); Data Curation (supporting); Writing - Original Draft (supporting). 
{\bf Michael Wand:}
Conceptualization (supporting); Methodology (supporting); Data Curation (supporting); Supervision (equal); Writing - Review \& Editing (supporting); Funding Acquisition (equal).
{\bf Karin Everschor-Sitte:}
Conceptualization (equal); Writing - Original Draft (supporting); Writing - Review \& Editing (supporting); Supervision (equal); Funding Acquisition (equal).
{\bf Friederike Schmid:}
Conceptualization (equal); Methodology (equal); Resources (lead); Writing - Review \& Editing (lead); Supervision (equal); Funding Acquisition (equal).

\medskip

\section{Data Availability}

The data that support the findings of this study and
the scripts used to generate them are openly available in
Zenodo (10.5281/zenodo.18256850) and GitHub (\url{https://github.com/ky-klos/full_chaincomplex_topological_data_analysis.git}, \url{https://github.com/ky-klos/cwgan_topological_defects.git})

\appendix

\section{Agreement Scores}

The agreement scores in the network validation discussion are calculated in
the following ways, where the superscripts $r$ and $g$ stand for
real/simulated and network generated data, respectively: 

For the physical measures $\chi$, $\Upsilon$, $C$, and $U$
in Fig.\ ~\ref{fig:main_physical_analysis}, 
we use normalized rooted mean squared error nRMSE, calculated by:
\begin{equation}
  \text{nRMSE} = 
     1 - \frac{\sqrt{\frac{1}{N}\sum^N_{i=0} (y^{r}_i 
         - y^{g}_i)^2}}{|\max(y^{r}) - \min(y^{r})|}.
  \label{equ:acc_observables}
\end{equation}
For the distributions of the energy $E$ and the
magnetization $M$, and the distributions entering the
topological data analysis in Fig.~\ref{fig:TDA_main}, we use
\begin{align}
  \tilde{W}_1(P^{r},P^{g}) &= 
     1. - \frac{W_1(P^{r},P^{g})}{\max(P^{r,g})-\min(P^{r,g})} \label{equ:acc_W1}\\
       W_1(P^{r},P^{g}) &= 
         \inf_\pi (\frac{1}{N}\sum^N_{i=1}||P^{r}_i -
         P^{g}_{\pi(i)}||). \label{equ:W1}
\end{align}


\bibliography{manuscript}
\resetpages          

\bibliographystyle{ieeetr}

\setlength{\fboxsep}{0pt}
\setlength{\fboxrule}{.1pt}

\title{Supplementary Material:
Reconstruction of spin structures from topological charge distributions via generative neural network systems}


\author{Kyra H. M. Klos}
\affiliation{%
 Institute of Physics, Johannes Gutenberg-University Mainz, 55128 Mainz, Germany
}%

\author{Jan Disselhoff}
\affiliation{%
 Institute of Computer Science, Johannes Gutenberg-University Mainz, 55128 Mainz, Germany
}%

\author{Michael Wand}
\affiliation{%
 Institute of Computer Science, Johannes Gutenberg-University Mainz, 55128 Mainz, Germany
}%

\author{Karin Everschor-Sitte}%
\affiliation{%
 Faculty of Physics and Center for Nanointegration Duisburg-Essen (CENIDE), University of Duisburg-Essen, 47057 Duisburg, Germany
}%

\author{Friederike Schmid}
\affiliation{%
 Institute of Physics, Johannes Gutenberg-University Mainz, 55128 Mainz, Germany
}%
\date{\today}

\maketitle

\onecolumngrid

\noindent
{\large \bf Additional Figures}

\bigskip
\bigskip

\begin{figure}[ht!]
\centering
\includegraphics[width=0.45\textwidth]{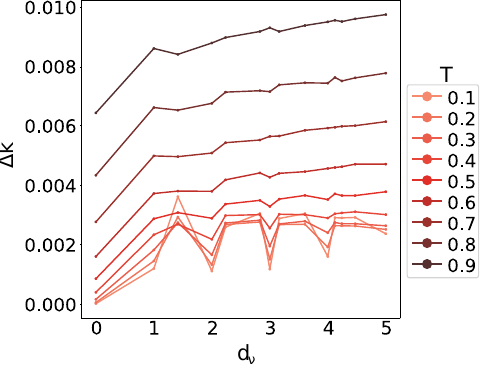}
  \caption{
 Mean difference $\Delta k = \frac{1}{N}\sum_r |\tilde{k}_r - k_r|$ 
 between true winding number $k_r$ and the approximate
 smooth version $\tilde{k}_r$ used in the loss function
 (main article, Eq.~(19))
 vs. defect pair distance $d_\nu$ at different temperatures 
 $k_B T/J \in [0.1:0.9]$  as indicated.
\label{fig:appendix_vortex_loss} }
\end{figure}

\begin{figure}[ht!]
\centering
\includegraphics[width=0.45\textwidth]{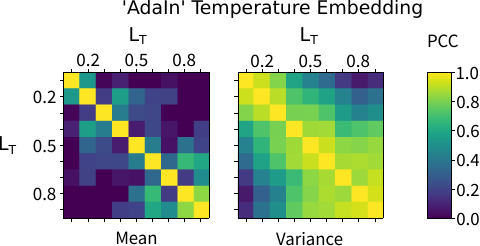}
\caption{Pearson correlation coefficient (PCC) matrix
 capturing the linear correlation between the means and variances of
 the last layer of the trained 'AdaIn' normalization for the
 temperature label embedding $L_T$ with temperatures $k_B T/J \in [0.1,
 0.9]$).
\label{fig:adain_correlation} }
\end{figure}

\begin{figure}[ht!]
\centering
\includegraphics[width=\textwidth]{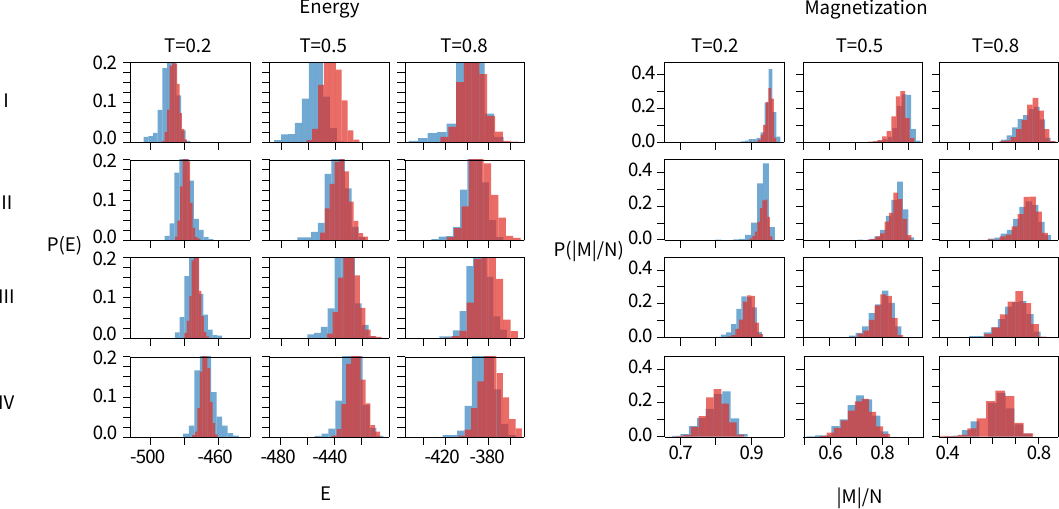}
  \caption{
  Histograms of energy $E$ (left) and 
  magnetization per spin $|M|/N$ (right) obtained from 1000 samples of
  simulated (red) and generated (blue) spin configurations 
  with  defect pair distances $d_{\nu} = 0.0,1.0,3.0,5.0$ (I-IV) and 
  temperatures $T = 0.2 J/k_B, \: T = 0.5 J/k_B, \: T = 0.8 J/k_B$
  as indicated.
\label{fig:distribution_energy_magnetization} }
\end{figure}

\begin{figure*}
\centering
  \hspace{-1.0cm}
  \includegraphics[width=0.9\textwidth]{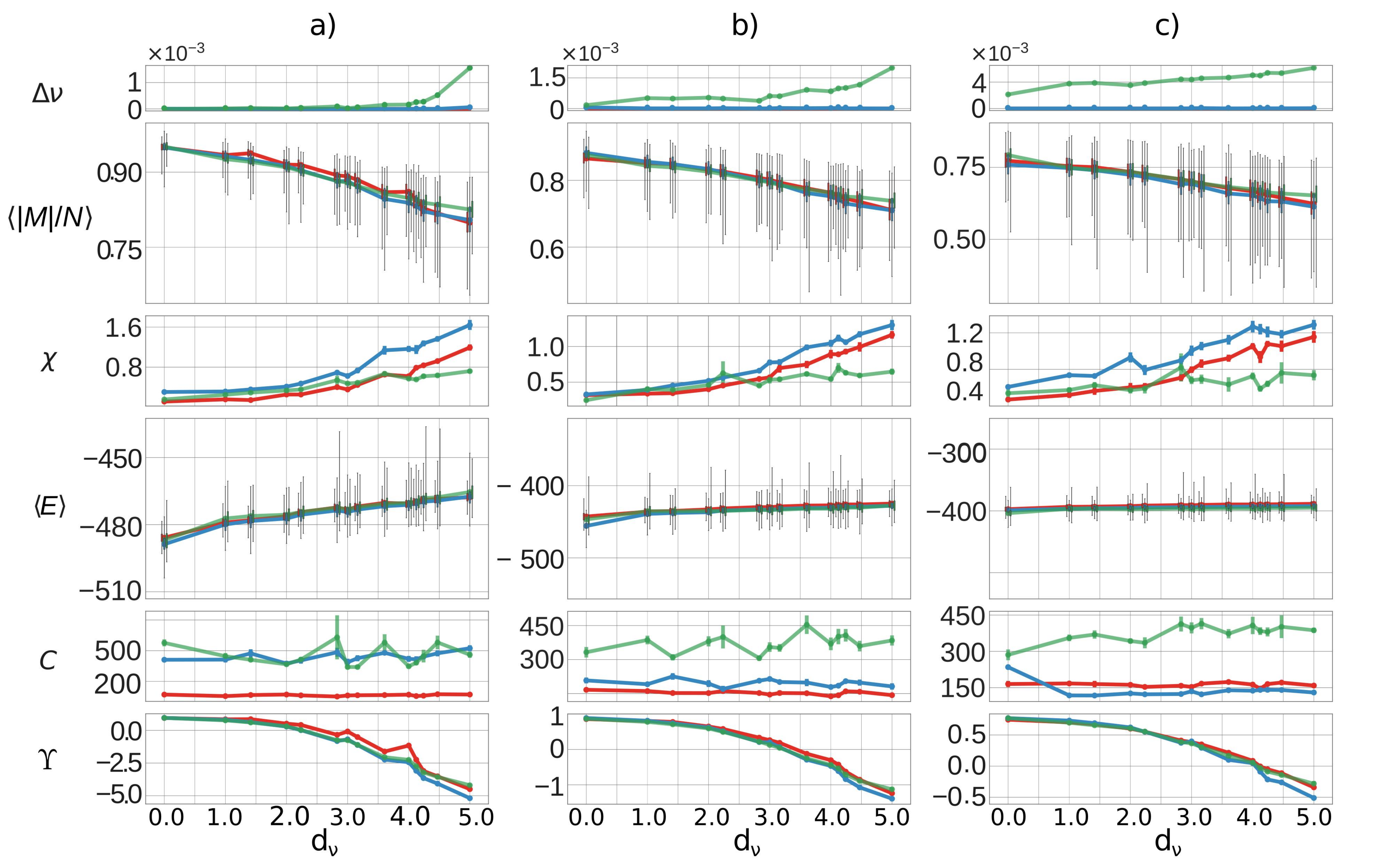}
 \vspace*{-1\baselineskip}
  \caption{Impact of Fourier critic on the quality of generated 
  spin configurations. The graphs compare physical observables obtained from
 configurations generated by the full WGAN presented in the main article
 (blue), by a simpler WGAN that does not include a separate Fourier
 critic (green), and from simulated configurations (red), plotted versus
 mean defect pair distance $d_{\nu}$ for a) $T=0.2 \, J/k_B$, b) $T=0.5
 \, J/k_B$, and c) $T=0.8 \, J/k_B$. The sample size is 1000 per
 parameter combination.  The observables are: defect configuration
 $\Delta\nu$ (main article, Eq.~(23)), mean magnetization per spin,
 $\langle|M|/N\rangle$, magnetic susceptibility $\chi$ (main article,
 Eq.~(5)), mean energy $\langle E \rangle$, specific heat $C$  (main
 article, Eq.~(6)) and helicity modulus $\Upsilon$ (main article,
 Eq.~(4)). Compared to the full WGAN (main text), training of the
 simpler WGAN involved changing the vortex regularization factor
 to $\beta=50$ and an additional instance normalization in the
 critic architecture. Apart from that, the training process and 
 the architectures of generator and critic were identical. 
 }
\label{fig:physical_analysis_compare_wo_fourier}
\end{figure*}

\begin{figure*}
\centering
  \hspace{-1.0cm}
  \includegraphics[width=0.5\textwidth]{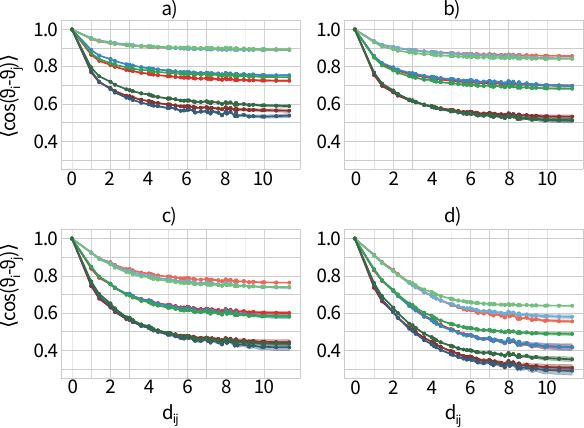}
 \vspace*{-1\baselineskip}
  \caption{Impact of Fourier critic on the quality of generated
  spin configurations. The graphs compare spin pair correlation functions 
   in configurations generated by the full WGAN presented in the main article
 (blue), by a simpler WGAN that does not include a separate Fourier
 critic (green), and in simulated configurations (red),
 at temperatures $k_B T/J \in [0.2, 0.5, 0.8]$ (encoded by lightest
 to darkest color) and for defect distances a) $d_\nu=0$,
 b) $d_\nu = 1$, c) $d_\nu = 3$, $d_\nu = 5$.
\label{fig:correlationfunction_compare_wo_fourier}
}

\end{figure*}

\begin{figure*}
\centering
\includegraphics[width=0.95\textwidth]{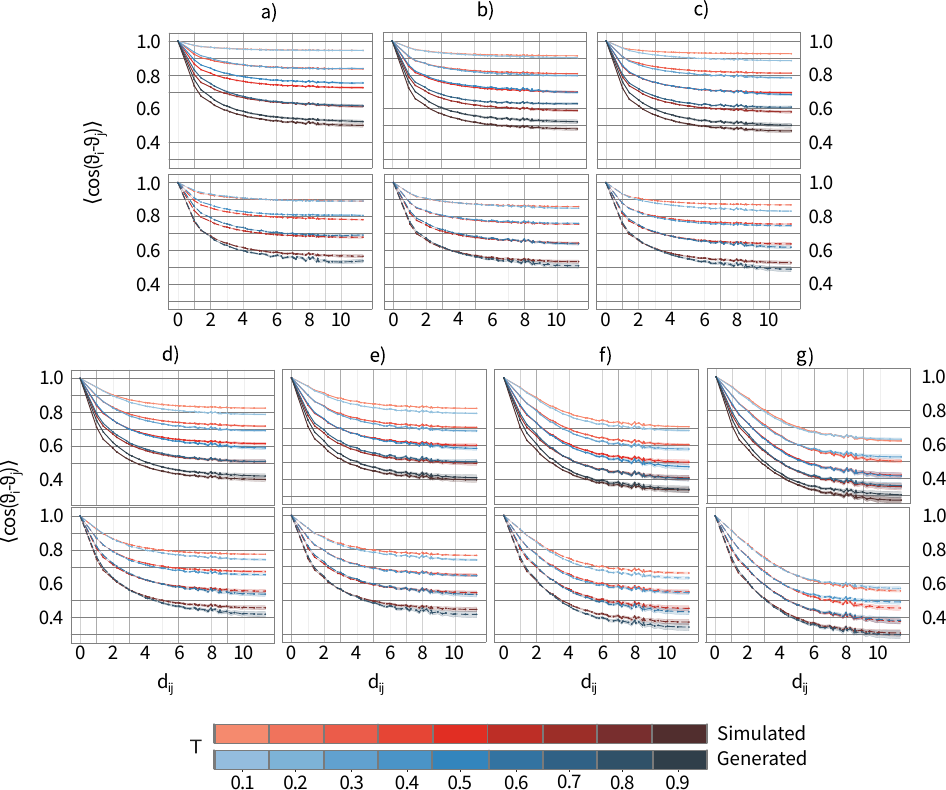}
  \caption{Spin-spin correlation functions in the full temperature
  range for defect field distances \\$d_{\nu} \in
  [0.0,1.0,\sqrt{2},2\sqrt{2},3.0,4\sqrt{2},5.0] $ shown in a)-g)
  respectively with odd temperatures on top and even temperatures at
  bottom- separated for visibility).  Here (blue) represent generated
  neural network produced data and (red) the simulated test data,
  ordered by temperature according to the color scheme at the bottom.
\label{fig:appendix_correlation} }
\end{figure*}

\begin{figure*}
\centering
\includegraphics[width=0.9\textwidth]{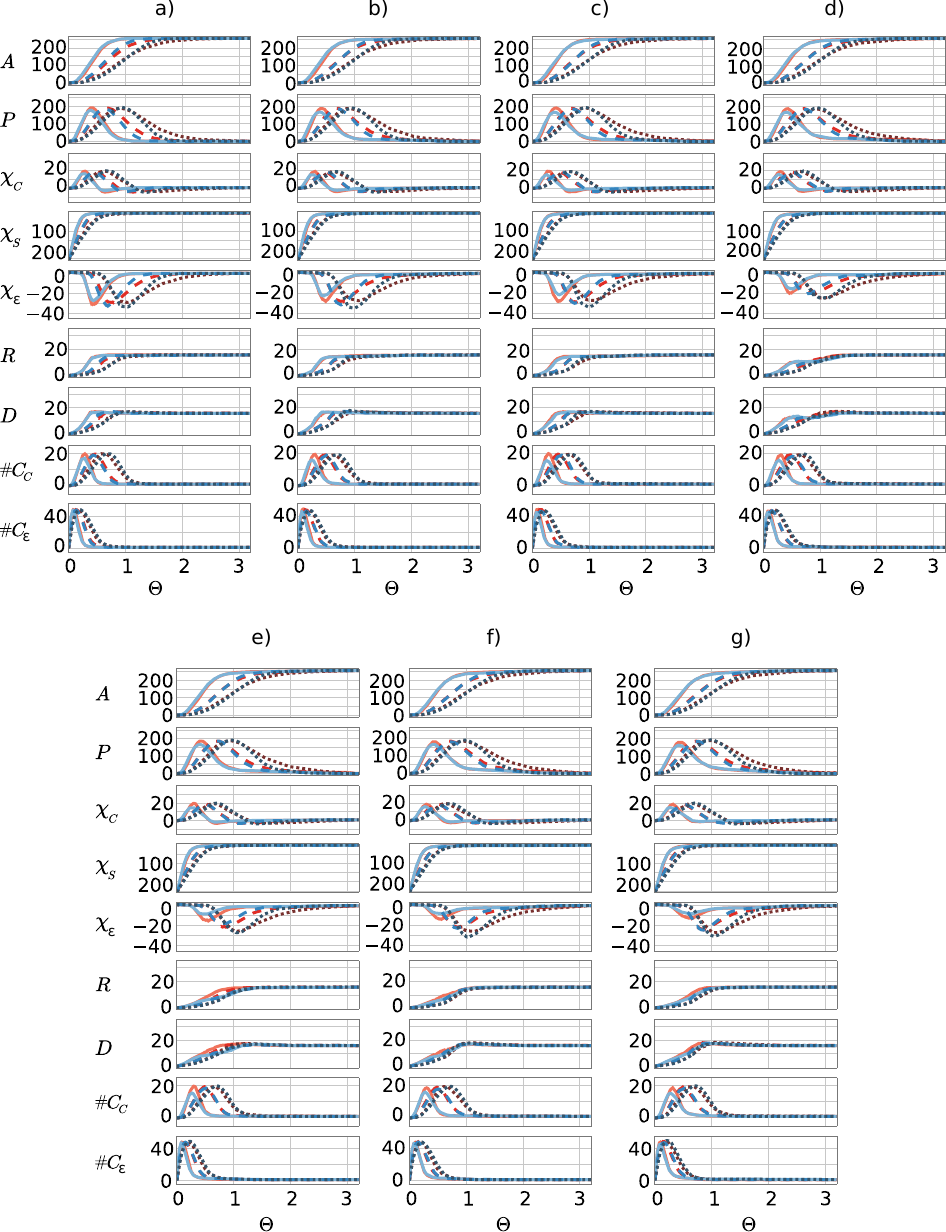}
\caption{Geometrical and topological measures as described in Section
    {\em Topological data analysis} versus filtration parameter $\Theta$,
    calculated from generated (blue) and simulated (red) data for
    representative temperatures $T\in[0.2,0.5,0.8] J/k_B$ (light solid,
    dashed, and dark dotted, respectively) and defect distances $d_{\nu} \in
    [0.0,1.0,\sqrt{2},2\sqrt{2},3.0,4\sqrt{2},5.0] $ from (a)-(g). 
    Specifically, the graphs shows the Minkowski
    measures of the plaquette subset ${\cal C}_\Theta$, i.e.,
    area ($A$), perimeter ($P$), and Euler characteristics
    $\chi_{\cal C}$ 
    (main article, Eq.~(12));
    The Euler characteristics derived from the Betty numbers
    $\chi_{\cal S}$ 
    (Eq.~(11));
    The mean Euler characteristics of connected components
    of ${\cal E}_\Theta \cup {\cal C}_\Theta$
    $\chi_{\cal E}$ 
    (Eq.~(15));
    The corresponding graph radius 
    (Eq.~(14))
    and diameter 
    (Eq.~(13)), 
    and
    the average number of plaquette clusters $\#C_{\cal C}$ and
    edge clusters $\# C_{\cal E}$.
     \label{fig:appendix_TDA} }
\end{figure*}

\begin{figure*}
\centering
  \hspace{-1.0cm}
  \includegraphics[width=0.9\textwidth]{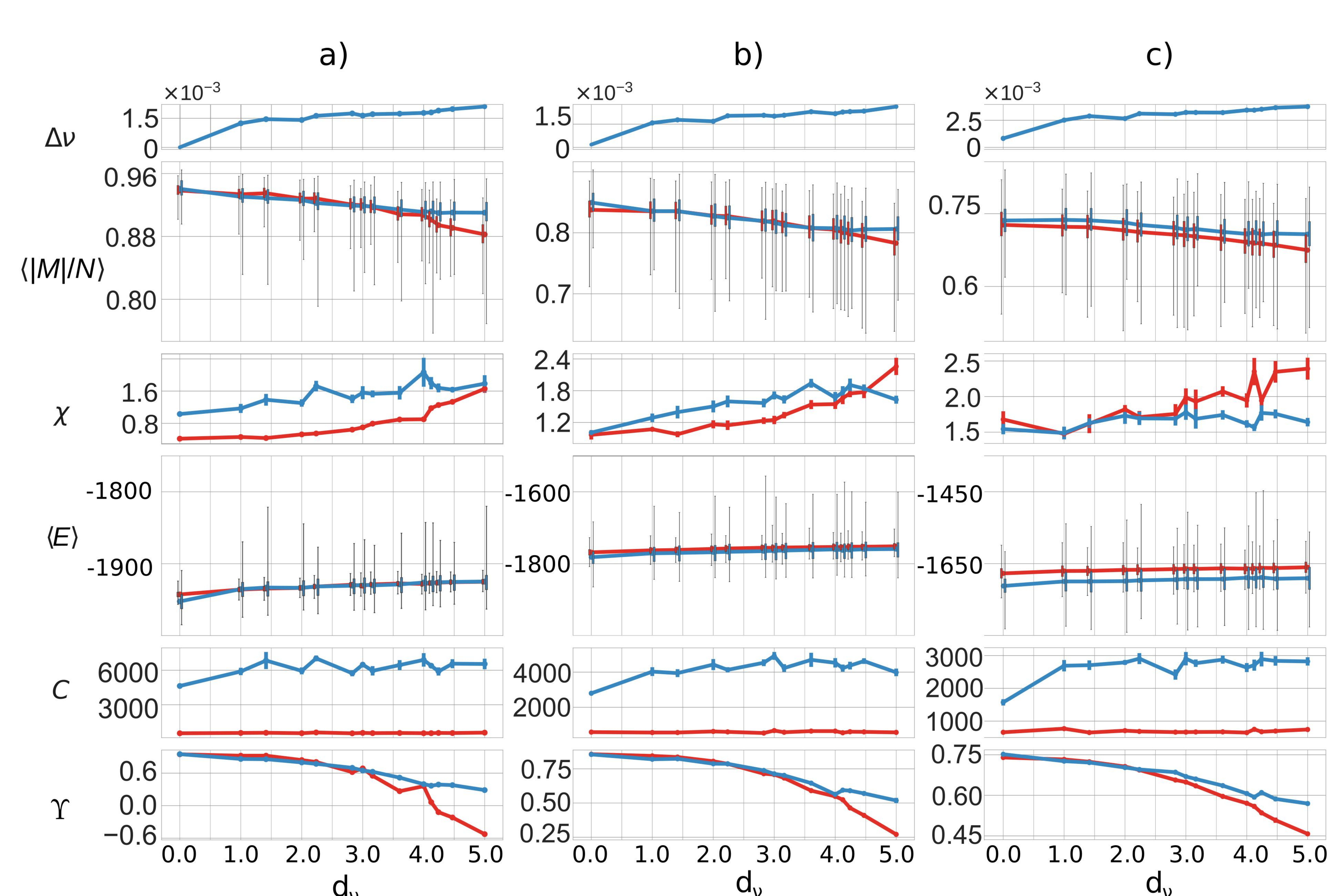}
 \vspace*{-1\baselineskip}
  \caption{Scalability of the network architecture with 
  respect to system size. The graphs show
  physical observables versus mean defect pair distance
  $d_{\nu}$ for $T=0.2 \, J/k_B$ (a), $T=0.5 \, J/k_B $ (b) and 
  $T=0.8 \, J/k_B$ (c) in systems with lattice size $L=32$.
  Data were abtained by analysis of 1000 simulated (red) and
  generated (blue) spin lattice configurations per parameter
  combination after retraining the network for lattice size $L=32$.
  The observables are: Defect configuration difference $\Delta\nu$
  Defect configuration difference $\Delta\nu$ 
  (main article, Eq.~(23)), mean magnetization per spin, 
  $\langle|M|/N\rangle$, magnetic susceptibility
  $\chi$ (main article, Eq.~(5)), mean energy $\langle E \rangle$,
  specific heat $C$  (main article, Eq.~(6)) and helicity modulus
  $\Upsilon$ (main article, Eq.~(4)). Red lines correspond to
  data from simulations, blue lines to data from generated
  configurations. The generated data here is produced by the 
  WGAN architecture presented in the main article,
  which was retrained with reduced generator learning rate 
  $5\times10^{-5}$ on data with $N_{d} \in [0,2]$ and
  lattice size $L=32$.
  }

\label{fig:physical_analysis_scaling_32}
\end{figure*}
\begin{figure*}
\centering
 \includegraphics[width=0.45\textwidth]{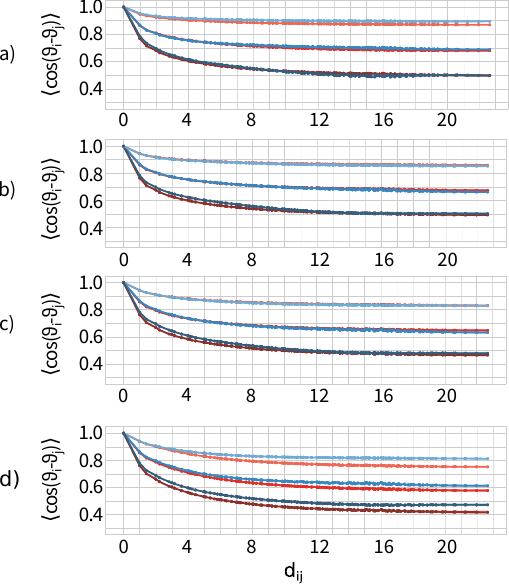}
 \vspace*{-1\baselineskip}

  \caption{Scalability of the network architecture with respect
  to system size. The graphs show
  spin-spin correlation functions from simulation data
 (red lines) and generated configurations (blue) in systems of size
 $L=32$ for defect distances $d_{\nu} \in [0.0,1.0,3.0,5.0] $ 
 (a)-d)) and temperatures $T\in[0.2,0.5,0.8]$ (lightest to
 darkest color.) The network architecture is that described
 in the main article, retrained with reduced generator learning rate
 $5\times10^{-5}$ on data with $N_{d} \in [0,2]$ and lattice size
 $L=32$.
 }

\label{fig:correlationfunction_scaling_32}
\end{figure*}

\begin{figure*}
\centering
  \hspace{-1.0cm}
  \includegraphics[width=\textwidth]{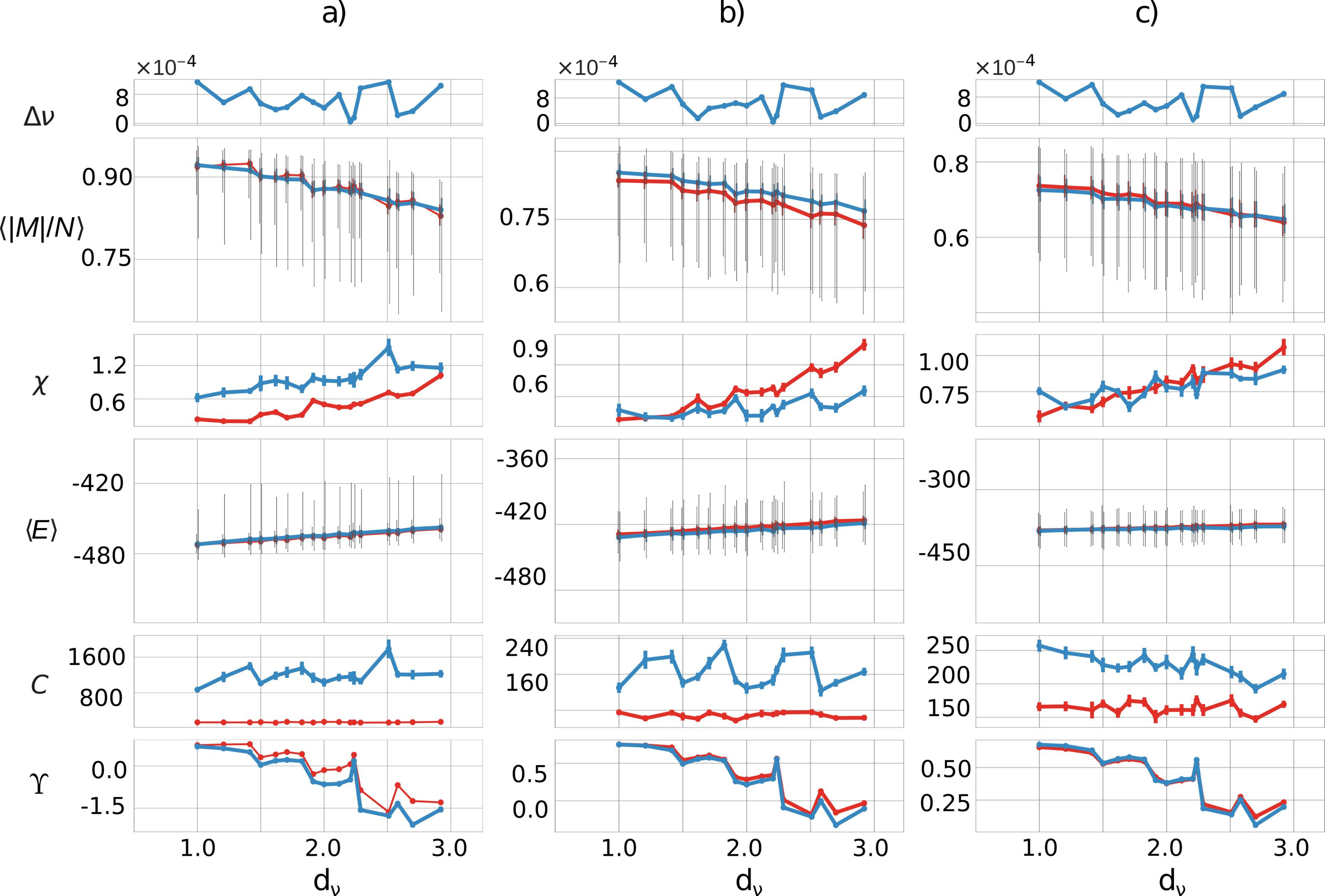}
 \vspace*{-1\baselineskip}
  \caption{Scalability of the network architecture with respect
  to defect number. The graphs show physical observables versus mean 
  defect pair distance $d_{\nu}$ for a) $T=0.2 \, J/k_B$, b) $T=0.5 \, J/k_B$, 
 and c) $T=0.8 \, J/k_B$ in systems with $L=16$ and $N_d=4$ defects. Data were
 obtained by analysis of 1000 simulated (red) and generated (blue)
 spin lattice configurations per parameter combination after retraining
 the network with $N_d=4$.  The observables are: defect configuration
 difference $\Delta\nu$ (main article, Eq.~(23)), mean magnetization per 
 spin $\langle|M|/N\rangle$, magnetic susceptibility $\chi$ (main article,
 Eq.~(5)), mean energy $\langle E \rangle$, specific heat $C$  (main
 article, Eq.~(6)) and helicity modulus $\Upsilon$ (main article,
 Eq.~(4)). The generated data were produced by the WGAN
 architecture presented in the main article, after retraining
 with reduced generator learning rate $5 \times 10^{-5}$ on data
 with $N_d \in [0,4]$ and $L=16$.
 }
\label{fig:physical_analysis_scaling_4}
\end{figure*}

\begin{figure*}
\centering
  \hspace{-1.0cm}
  \includegraphics[width=\textwidth]{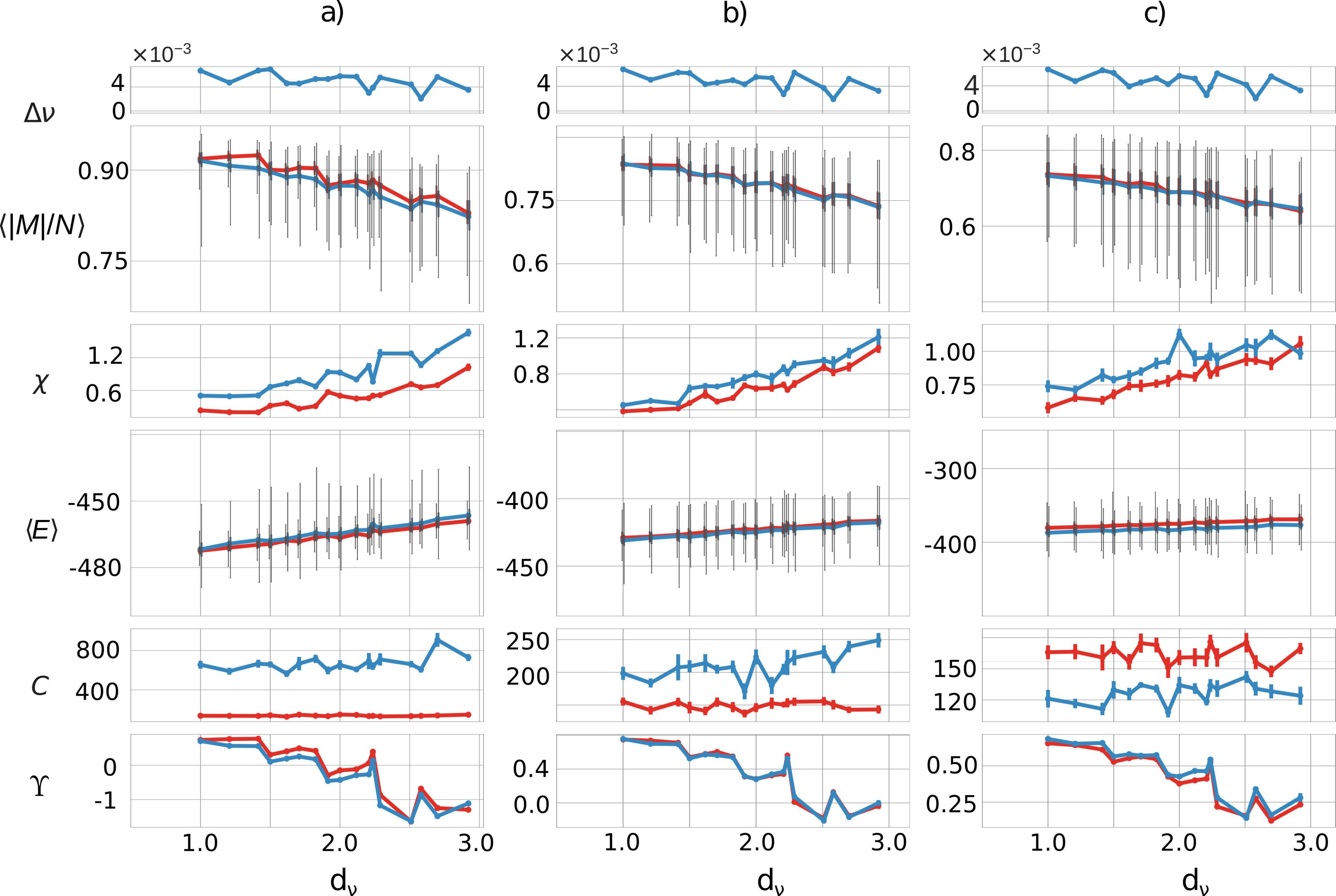}
 \vspace*{-1\baselineskip}
  \caption{Transferability of the network architecture to higher
  defect numbers.  The graphs show physical observables versus mean 
  defect pair distance $d_{\nu}$ for a) $T=0.2 \, J/k_B$, 
  b) $T=0.5 \, J/k_B$,
 and c) $T=0.8 \, J/k_B$ in systems with and $N_d=4$ defects.
 Data were obtained by analysis of 1000 simulated (red) and 
 generated (blue) spin lattice configurations per parameter 
 combination using the original WGAN network (main article) trained on data
 with $N_d \in [0:2]$.  The observables are: Defect configuration 
 $\Delta\nu$ (main article, Eq.~(23)), mean magnetization per spin, 
 $\langle|M|/N\rangle$, magnetic susceptibility
 $\chi$ (main article, Eq.~(5)), mean energy $\langle E \rangle$,
 specific heat $C$  (main article, Eq.~(6)) and helicity modulus
 $\Upsilon$ (main article, Eq.~(4)).}

\label{fig:physical_analysis_generalization_4_defects}
\end{figure*}

\begin{figure*}
\centering
  \hspace{-1.0cm}
  \includegraphics[width=\textwidth]{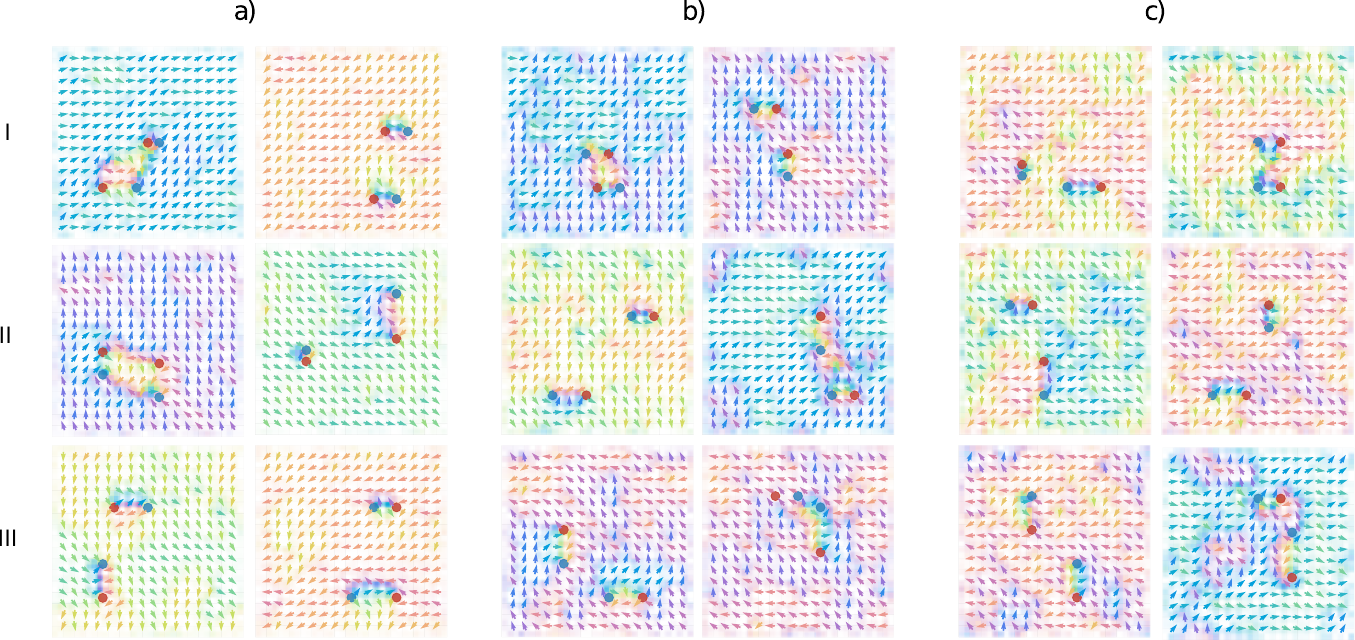}
 \vspace*{-1\baselineskip}
  \caption{
  Exemplary spin configurations generated through generalization of the
 WGAN network presented in the main article, which was trained
 on data with zero or two defects. The images correspond to
 temperatures a) $k_B T/J = 0.2$, b) $k_B T/J = 0.5$, and
 c) $k_B T/J =0.8$ and mean defect pair distances of (I)
 (I) $d_{\nu}=2.$, (II) $d_{\nu}=2.5$, and (III) $d_{\nu}=3.0$.}
\label{fig:spin_images_generalization_4_defects}
\end{figure*}


\end{document}